\documentclass[preprint,review,12pt]{elsarticle}

\usepackage{amssymb}
\usepackage{amsmath}
\usepackage{booktabs}
\usepackage{float}

\journal{elsvier journal}

\begin{document}

\begin{frontmatter}



\title{A Generalized Density Dissipation for Weakly-compressible SPH}


\author[label1,label2]{B.X. Zheng}

\affiliation[label1]{organization={Guangdong Provincial Key Laboratory of Turbulence Research and Applications, Department of Mechanics and Aerospace Engineering, Southern University of Science and Technology},
            city={ShengZhen},
            postcode={518055}, 
            country={China}}

\affiliation[label2]{organization={Department of Mathematics, University of Oslo},
            city={Oslo},
            postcode={0851}, 
            country={Norway}}

\author[label3]{Z.W. Cai}

\affiliation[label3]{organization={China Ship Scientific Research Center},
            city={Wuxi},
            postcode={214082}, 
            country={China}}

\author[label4,label5]{P.D. Zhao}

\affiliation[label4]{organization={School of Mechanical Engineering, Shanghai Jiao Tong University},
            city={Shanghai},
            postcode={200240}, 
            country={China}}
\affiliation[label5]{organization={School of Naval Architecture and Ocean Engineering, Dalian University of Technology},
            city={Dalian},
            postcode={116024}, 
            country={China}}

\author[label6]{X.Y. Xu}

\affiliation[label6]{organization={School of Computer Science and Technology, Xi'an University of Science and Technology},
            city={Xi’an},
            postcode={710054}, 
            country={China}}

\author[label2]{T.S. Chan\corref{cor1}}
\ead{taksc@uio.no}

\author[label1,label7]{P. Yu\corref{cor1}}
\ead{yup6@sustech.edu.cn}

\affiliation[label7]{organization={Center for Complex Flows and Soft Matter Research, Southern University of Science and Technology},
            city={ShenZhen},
            postcode={518055}, 
            country={China}}
            
\cortext[cor1]{Corresponding author}

\label{abstract}
\begin{abstract}
The weakly compressible Smoothed Particle Hydrodynamics (SPH) is known to suffer from the pressure oscillation, which would undermine the simulation stability and accuracy. To address this issue, we propose a generalized density dissipation scheme suitable for both single-phase and multiphase flow simulations. Our approach consists of two components. Firstly, we replace the basic density dissipation with the density increment dissipation to enable numerical dissipation crossing the interfaces of different fluids in multiphase flow. Secondly, based on the dissipation volume conservation, we utilize dissipation volume correction factor (VCF) to stabilize the simulations for multiphase flows with large density ratio. We demonstrate the accuracy, stability, and robustness of our method through four three-dimensional benchmarks, i.e., the sloshing under external excitations, the single and double bubbles rising, Rayleigh-Taylor instability, and Kelvin-Helmholtz instability. Additionally, our study reveals the relationship between SPH with the density dissipation and the approximate Riemann solver.

\end{abstract}

\begin{graphicalabstract}
\includegraphics[width=1\textwidth]{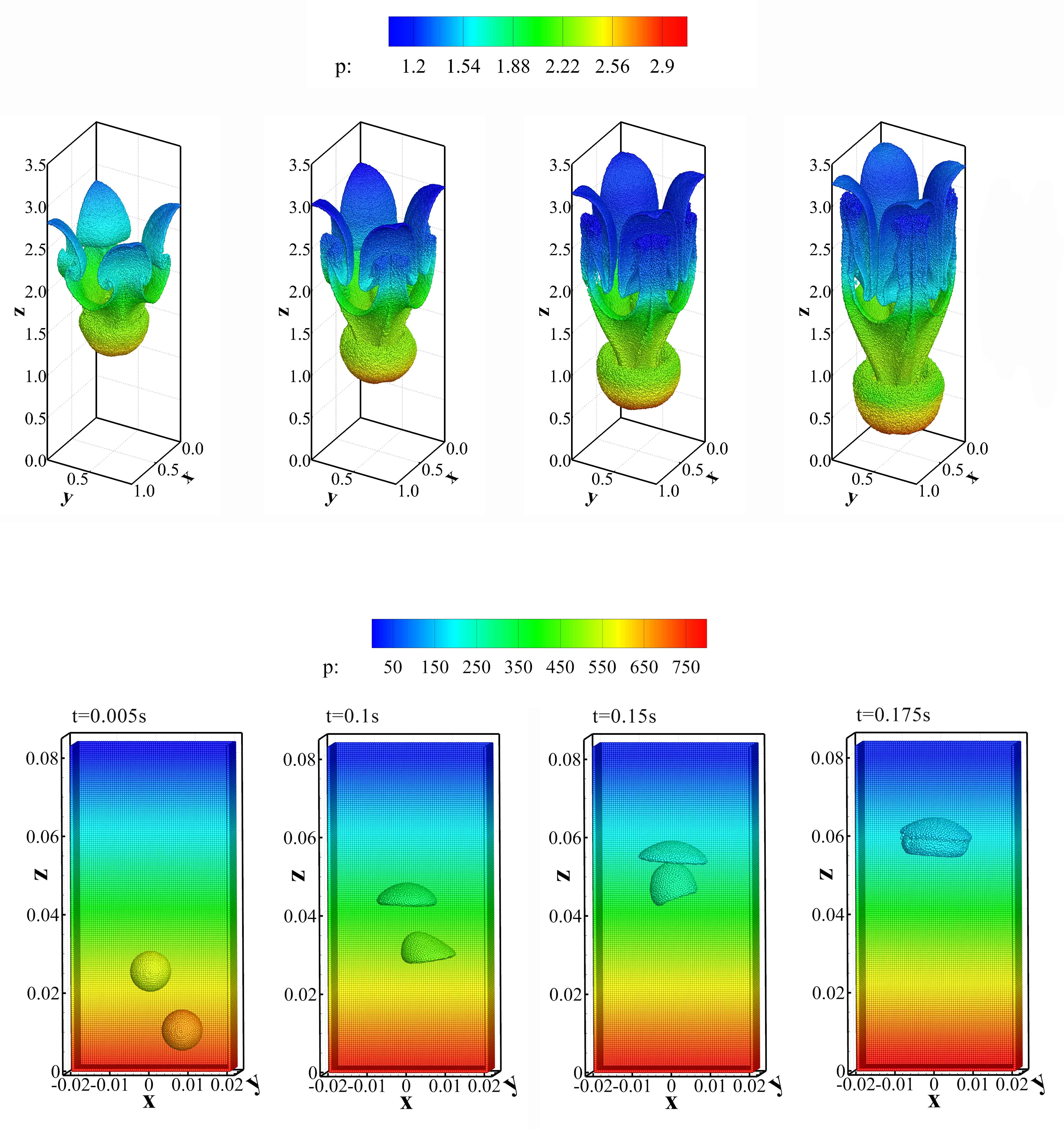}
\end{graphicalabstract}

\begin{highlights}
\item A generalized dissipation is proposed to suppress pressure oscillations in SPH 
\item The present dissipation scheme can cross different phases in multiphase flow
\item Relationship between dissipation and approximate Riemann solver is revealed
\item Four 3D benchmarks illustrate the capability of the present SPH method 
\end{highlights} 


\end{frontmatter}


\section{Introduction}
Smoothed Particle Hydrodynamics (SPH) is a widely recognized and purely mesh-less numerical method used for flow simulation. The method discretizes the fluid domain into finite fluid particles and then solves the unknown physical variables of each particle, including velocity, position, density, pressure, etc., to approximate the entire flow field. Over the past decades, SPH has proven to be highly effective in handling complex variables in flow problems, including fluid-structure interactions \cite{liu2019smoothed,sun2021accurate,zhang2021multi,zhang2017smoothed},  free-surface and multiphase flow \cite{monaghan1994simulating,colagrossi2003numerical,HU2006844,chen2015sph, xu2020modeling}, turbulence \cite{monaghan2002sph,antuono2021smoothed,shao2006incompressible}, etc.

The weakly compressible SPH is a common variation of the SPH method for simulating incompressible flow. It introduces a numerical sound velocity to establish the relationship between pressure and particle density, allowing for the explicit time-stepping method to update the velocity, position, and density at each time step, and thus eliminating the need to solve the large-scale matrix equation every time, as required in the semi-implicit method of incompressible SPH \cite{hughes2010comparison}. Additionally, weakly compressible schemes are generally better suited for modeling free-surface flows as the boundary condition along the free surface is implicitly satisfied \cite{antuono2012numerical}. However, weakly compressible SPH is known to suffer from pressure oscillation, which can significantly affect the stability of the numerical simulation and the accuracy of the simulated results.

To address pressure oscillation in weakly compressible SPH, various pressure correction algorithms have been proposed. In weakly compressible SPH, the fluid particle pressure is calculated based on the updated density of fluid particles. Therefore, pressure correction algorithms are implemented by modifying the density of particles. Colagrossi and Landrini \cite{colagrossi2003numerical} developed a method that uses the Moving-Least-Squares (MLS) interpolation proposed by Dilts \cite{dilts1999moving} to re-initialize the density field, obtaining a more regular pressure distribution. Molteni et al. \cite{molteni2009simple} proposed a convenient and effective pressure correction method by inserting a pure numerical density dissipation term in the continuous equation. The SPH model with the similar numerical density dissipation term, called $\delta$-SPH, has been developed in recent years. Antuono et al. \cite{antuono2010free} extended the pressure dissipation term to a higher order and successfully applied it to the simulation of free surface flow. Sun et al. \cite{sun2017deltaplus,sun2019consistent} incorporated particle shifting technique (PST) into the $\delta$-SPH and proposed $\delta$-plus-SPH, which demonstrated very good stability in long-term simulations of both two-dimensional (2D) and three-dimensional (3D) flow problems.

It is worth noting that both the MLS method and the density dissipation of the $\delta$-SPH cannot be directly applied at the interface of the multi-phase flow due to the discontinuity of the fluid density at the interface. As a result, when utilizing these density dissipations in multiphase flow simulations, they can only be applied independently for each phase. Mokos \cite{mokos2017multi} applied the MLS method to the pressure correction of air and water respectively when simulating the two-phase violent flow problem. Hammani \cite{hammani2020detailed} also used this phase-separated pressure correction to implement $\delta$ dissipation to multiphase flow simulations. Although it is not difficult for particle methods to distinguish particles of different phases, we naturally expect a more generalized and intrinsic dissipation scheme that is equally valid at interfaces.

Besides the density correction algorithm, numerical schemes such as SPH with an approximate Riemann solver  can also stabilize the pressure field and be directly applied to multiphase flow simulation.  Monaghan \cite{monaghan1997sph} firstly pointed out that the SPH momentum and thermal energy equations with  artificial dissipation are very similar to the equations constructed for Riemann solutions of compressible gas dynamics. Puri and Ramachandran \cite{puri2014approximate} showed the equivalence between the dissipative terms of Godunov SPH (GSPH) and the signal based SPH artificial viscosity, under the restriction of a class of approximate Riemann solvers. However, since neither of the above two studies focused on the pressure field stability in the flow problem, only the dissipation term in the momentum equation was discussed, but not the dissipation in the continuity equation. Recently, researchers applied SPH with the approximate Riemann solvers to simulate flow problem and obtained accuracy and stable results for pressure field. Zhang et al. \cite{zhang2017weakly} presented a weakly-compressible SPH with a modified Riemann solver to predict the single-phase free-surface flow. In their work, a stable pressure field was obtained when simulating the impacting flow problems. Meng et al. \cite{meng2020multiphase} proposed a SPH model based on Roe’s approximate Riemann solve. Utilizing their model, stable and smooth interfacial pressure can be obtained even for flows with high density ratios. Other successful applications of SPH in combination with approximate Riemann solutions in multiphase flow simulations demonstrated the effectiveness of such methods in stabilizing the pressure field \cite{meng2021numerical,fang2022accurate,hammani2020detailed}. However, the above studies did not explore the underlying mechanism on why SPH combined with the approximate Riemann solution can stabilize the pressure field for both single-phase and multi-phase flow simulations.

To shed some light on the above mentioned matters, we propose a generalized density dissipation term for SPH that effectively suppresses pressure oscillations in both multiphase and free-surface flow simulations in this work. The implementation of this dissipation consists of two parts: firstly, the dissipation of density increments is utilized instead of density dissipation, which allows numerical dissipation to cross interfaces of different fluids. Secondly, we incorporate dissipation volume correction factor, based on the dissipation volume conservation, to ensure stable simulated results in large-density-ratio problems. Furthermore, by revealing the relationship between between SPH combined the approximate Riemann solution and the present generalized density dissipation, we explain why SPH combined with the approximate Riemann solution can stabilize the pressure field for both single-phase and multi-phase flow simulations.   

The remaining parts of the paper are organized as follows. Section 2 presents the governing equations for the flow simulation and the SPH discretization. Section 3 details the algorithms of the present generalized density dissipation and discusses the relationship between the SPH with the present dissipation and the approximate Riemann solver. In Section 4, four 3D representative numerical examples, including sloshing under external excitations, bubble rising, Rayleigh-Taylor instability and Kelvin-Helmholtz instability are simulated by the present method for evaluation of its accuracy and stability. Conclusions are finally drawn in Section 5.

\section{Governing equations and SPH discretization}
\subsection{Governing equations}
The continuity and momentum equations to describe weakly compressible viscous flows under the Lagrangian coordinate are as follows \cite{zhang2015sph}:

\begin{equation}
\frac{\mathrm{d} \rho}{\mathrm{d}t}=-\rho \nabla \cdot  \boldsymbol{v}
\end{equation}

\begin{equation}
\rho \frac{\mathrm{d} \boldsymbol{v}}{\mathrm{d}t}=-\nabla P+\boldsymbol{F}^V+\boldsymbol{F}^B+\boldsymbol{F}^S
\end{equation}
where $\rho$, $\boldsymbol{v}$ and $P$ represent the density, the velocity and the pressure, respectively, and $\boldsymbol{F}^V$, $\boldsymbol{F}^B$, and $\boldsymbol{F}^S$ represent the viscous force per unit volume, the external body forces per unit volume, and the surface tension force per unit volume, respectively.

For an incompressible Newtonian fluid, the viscous force $\boldsymbol{F}^V$  can be simplified to \cite{HU2006844}
\begin{equation}
\boldsymbol{F}^V=\mu \nabla^2 \boldsymbol{v}
\end{equation}
where $\mu$ denotes the dynamic viscosity.

According to the continuum surface force (CSF) model \cite{brackbill1992continuum}, the surface tension force is considered as the body force, which gives
\begin{equation}
\boldsymbol{F}^S=-\sigma \kappa \hat{\boldsymbol{n}} \gamma_e
\end{equation}
where $\sigma$ is the surface tension coefficient, $\kappa$ is the interface curvature, $\hat{\boldsymbol{n}}$ is the unit normal vector, and $\gamma_e$ denotes the surface delta function which is applied to transform the surface tension force per unit interfacial area to per unit volume.

By utilizing the assumption of weak compressibility, the tedious process of resolving the pressure Poisson equation (PPE) is avoided. Instead, the pressure of the particle is correlated to its density increment  through the equation of state (EoS). As suggested by Rezavand \cite{rezavand2020weakly} a simple linear EoS equation is adopted for all phases.
\begin{equation}
p=c_s{ }^2\left(\rho-\rho_0\right)
\end{equation}
where $\rho_0$ denotes the reference density of the fluid and $c_S$ is the numerical sound speed which is used to control the compressibility of fluids. The incompressible limit requires the variation of density to be within $1 \%$ \cite{colagrossi2003numerical}. To fulfill this condition, $c_S=\sqrt{\mathrm{d} p / \mathrm{d} \rho}$ has to be at least one order of magnitude greater than the maximum flow velocity:
\begin{equation}
c_s \geq 10 \max (|\boldsymbol{v}|)_{\Omega}
\end{equation}
where $\Omega$ represents the computational domain.

\subsection{SPH discretization}

There exist several discrete forms \cite{monaghan2013simple, grenier2009hamiltonian, hu2007incompressible, HU2006844,colagrossi2003numerical} for Eqs. (1) and (2) within the framework of SPH, each with its own advantages and disadvantages that depend on the specific hydrodynamic characteristics being modeled. In the case of Eq. (1), the most straightforward discretization is employed in the present work as:
\begin{equation}
\frac{\mathrm{d} \rho_i}{\mathrm{d} t}=-\rho_i \sum_j V_j\left(\boldsymbol{v}_j-\boldsymbol{v}_i\right) \nabla_i W_{ij}
\end{equation}
where the subscripts $i$ and $j$ denote the indices of the center and the neighboring particles, respectively, $V$ is the particle volume, $\nabla_i$ represents the gradient operator acting on the kernel function of the particle $i$, and $W_{ij}=W(\boldsymbol{r}_i-\boldsymbol{r}_j, h)$ denotes the kernel function. In this paper, we choose Gaussian kernel for all the numerical examples, which reads:
\begin{equation}
W(\boldsymbol{r}_i-\boldsymbol{r}_j, h)= \begin{cases}\frac{e^{-R^2}}{h^d \pi^{d / 2}}, & R \leq 3 \\ 0, & R>3 \end{cases}
\end{equation}
where $\boldsymbol{r}$ is the position vector of a particle, $h$ denotes the smooth length, $R=\left|\boldsymbol{r}_i-\boldsymbol{r}_j\right| / h$ is the relative distance, and $d$ is the space dimension. 

For the discretization of the pressure gradient in the momentum equation, we adopt the method proposed by Grenier \cite{grenier2009hamiltonian}, which successfully simulates the problem of multiphase flow with the free surface. The pressure gradient is discretized as:

\begin{equation}
\nabla p =\sum_j\left(\frac{p_i}{\Gamma_i}+\frac{p_j}{\Gamma_j}\right) \nabla_i W_{ij} V_j
\end{equation}
where $\Gamma_i={\sum_j W_{i j} V_j+W_i V_i}$ when $i$ is a fluid particle and $\Gamma_i=1$ when $i$ is a boundary particle.

Based on the inter-particle averaged shear tensor \cite{flekkoy2000foundations}, Hu and Adams \cite{HU2006844} suggested the following discretization for the viscous force in the multiphase modeling
\begin{equation}
\boldsymbol{F}_i^V=\frac{4 \mu_i \mu_j}{\mu_i+\mu_j} \sum_j \frac{\boldsymbol{r}_{ij} \cdot \nabla_i W_{i j}}{{r}_{ij}^2+\eta^2}\left(\boldsymbol{v}_i-\boldsymbol{v}_j\right) V_j
\end{equation}
where $\boldsymbol{r}_{i j}=\boldsymbol{r}_i-\boldsymbol{r}_j$, $r_{i j}=\left|\boldsymbol{r}_{i j}\right|$ and $\eta$ is a parameter that prevents zero denominator and $\eta$ is set to $0.1 h$ in the present study.

According to Adami \cite{adami2010new} and Zhang \cite{zhang2015sph}, the discretization for the surface tension can be divided in the following steps.

An index number $c_i^j$ for particle $i$ is firstly defined as,
\begin{equation}
c_i^j= \begin{cases}\frac{2 \rho_i}{\rho_i+\rho_j} & \text { if particle } i \text { and } j \text { belong to different phases } \\ 0 & \text { if particle } i \text { and } j \text { belong to the same phase. }\end{cases}
\end{equation}

Then the unit normal $\hat{n}$ can be obtained as follows
\begin{equation}
\hat{\boldsymbol{n}}_i=\frac{\boldsymbol{n}_i}{\left|\boldsymbol{n}_i\right|}=\frac{\nabla c_i}{\left|\nabla c_i\right|}
\end{equation}
where $\nabla c_i$ is approximated as
\begin{equation}
\nabla c_i=\sum_j c_i^j \nabla_i W_{i j} V_j
\end{equation}

Finally, the curvature of particle $i$, $\kappa_i$, is calculated as follows,
\begin{equation}
\kappa_i=-\left(\nabla \hat{\boldsymbol{n}}_i\right)=-d \frac{\sum_j\left(\hat{\boldsymbol{n}}_i-\varphi_i^j \hat{\boldsymbol{n}}_j\right) \nabla_i W_{i j} V_j}{\sum_j\left|\boldsymbol{r}_{i j}\right| \cdot\left|\nabla_i W_{i j}\right| V_j}
\end{equation}
where $\varphi_i^j$ is a coefficient to invert the direction of the unit normal $\boldsymbol{\hat{n}_j}$ in the case that particle $i$ and particle $j$ belong to different phases \cite{zhang2015sph}, which reads,
\begin{equation}
\varphi_i^j= \begin{cases}-1 & \text { if particles } i \text { and } j \text { belong to different phases } \\ 1 & \text { if particles } i \text { and } j \text { belong to the same phase}\end{cases}
\end{equation}

Now the surface tension force in Eq. (4) can be expressed as
\begin{equation}
\boldsymbol{F}_i^S=-\sigma \kappa_i \hat{\boldsymbol{n}}_i\left|\nabla c_i\right|
\end{equation}
where $\left|\nabla c_i\right|$ serves as the weight function which is consistent with the term of $\gamma_e$ in Eq. (4).

\subsection{Boundary conditions}

In SPH, a commonly used method for implementing boundaries of solid objects is to distribute dummy fluid particles. The number of the layers of boundary particles should ensure that the support domain of every fluid particle in the flow field is not truncated. The positions and velocities of the boundary particles are determined by the physical conditions of the boundaries. The pressure of a boundary particle are determined by,
\begin{equation}
p_i^{\text {boundary}}=\frac{\sum_j W_{i j} p_j+W_i p_i}{\sum_j W_{i j} V_j+W_i V_i}
\end{equation}  
considering the EoS equation, the density is calculated as,
\begin{equation}
\rho_i^{\text {boundary}}=\frac{p_i}{{c_s}^2}+\rho_{0max}
\end{equation}  
where $\rho_{0max}$ is the initial density of the boundary particles, which is equal to the initial density of the densest fluid particles in a multiphase flow. The volume of the boundary particles are also updated by considering 
\begin{equation}
V_i^{\text {boundary}}=\frac{m_i^{\text {boundary}}}{\rho_{i}^{\text {boundary}}}
\end{equation} 
where $m_i^{\text {boundary}}$ is specified based on the density, initial pressure, and initial volume during initialization and never changes during the calculation.

\subsection{Time marching scheme}
The evolutions of the fluid properties are achieved by choosing a proper time marching scheme to discretize the time derivative terms in the governing equations. The predictor-corrector scheme \cite{monaghan1989problem} with second-order accuracy is adopted in the present work, which gives:

(1) The predictor step:
\begin{equation}
\left\{\begin{array}{l}
\boldsymbol{v}_i^{n+1 / 2}=\boldsymbol{v}_i^n+\frac{\Delta t}{2}\left(\frac{\mathrm{d} \boldsymbol{v}}{\mathrm{d} t}\right)_i^n \\
\rho_i^{n+1 / 2}=\rho_i^n+\frac{\Delta t}{2}\left(\frac{\mathrm{d} \rho}{\mathrm{d} t}\right)_i^n \\
\boldsymbol{r}_i^{n+1 / 2}=\boldsymbol{r}_i^n+\frac{\Delta t}{2} \boldsymbol{v}_i^n
\end{array}\right.
\end{equation}
where the superscript $n$ denotes the number of the time steps and $\Delta t$ denotes the time step.

(2) The corrector step:
\begin{equation}
\left\{\begin{array}{l}
\boldsymbol{v}_i^{n+1 / 2}=v_i^n+\frac{\Delta t}{2}\left(\frac{\mathrm{d} v}{\mathrm{~d} t}\right)_i^{n+1 / 2} \\
\rho_i^{n+1 / 2}=\rho_i^n+\frac{\Delta t}{2}\left(\frac{\mathrm{d} \rho}{\mathrm{d} t}\right)_i^{n+1 / 2} \\
\boldsymbol{r}_i^{n+1 / 2}=\boldsymbol{r}_i^n+\frac{\Delta t}{2} \boldsymbol{v}_i^{n+1 / 2}
\end{array}\right.
\end{equation}

Finally, the fluid properties of the particles at the new time step are obtained as follows,
\begin{equation}
\left\{\begin{array}{l}
\boldsymbol{v}_i^{n+1}=2 \boldsymbol{v}_i^{n+1 / 2}-\boldsymbol{v}_i^n \\
\rho_i^{n+1}=2 \rho_i^{n+1 / 2}-\rho_i^n \\
\boldsymbol{r}_i^{n+1}=2 \boldsymbol{r}_i^{n+1 / 2}-\boldsymbol{r}_i^n
\end{array}\right.
\end{equation}

Considering the different CFL conditions based on the maximum artificial sound speed and the maximum flow speed, the surface tension, the body force, and the viscous force\cite{adami2010new,zhang2015sph},the final time step $\Delta t$ is constrained by
\begin{equation}
\Delta t \leq \min \left(\frac{0.25h}{c_s+\max (|v|)_{\Omega}},0.5\left(\frac{\rho h^3}{2 \pi \sigma}\right)^{\frac{1}{2}},0.25\left(\frac{h}{|g|}\right)^{\frac{1}{2}}, \frac{0.125\rho h^2}{\mu}\right)
\end{equation}

\section{The generalized density dissipation}
\subsection{Analysis of the basic density dissipation in $\delta$-SPH}
As mentioned in the introduction, the weakly-compressible SPH suffers from the pressure oscillation. An efficient way to solve this problem is to include the dissipation term on the right-hand side of the continuity equation as follows 
\begin{equation}
\frac{\mathrm{d} \rho}{\mathrm{d} t}=-\rho \nabla \cdot \boldsymbol{v}+\mathcal{D}
\end{equation}
where $\mathcal{D}$ is the dissipation term. Compared with the other two terms in Eq. (24), $\mathcal{D}$ must be small so that it does not affect the solution of the equation.

In the well-known $\delta$-SPH model \cite{molteni2009simple, antuono2010free, marrone2011delta, hammani2020detailed, sun2017deltaplus, antuono2012numerical}, the dissipation term is,
\begin{equation}
\mathcal{D}=\delta hc_s\nabla^2\rho
\end{equation}
where $\delta$ is a control parameter that adjusts the strength of dissipation.

Notice that $\nabla^2\rho$ in Eq. (25) is a second derivative term. Several schemes \cite{flebbe1994smoothed,takeda1994numerical,brookshaw1985method} has been proposed to discrete the second derivative. The scheme with the finite-difference-like form proposed by Brookshaw \cite{brookshaw1985method} is most recommended for the complicated flow simulation \cite{basa2009robustness,fatehi2011error}. By this means, for an arbitrary function $f$ defined at the position of the particle $i$, $\nabla^2f$ can be discretized as
\begin{equation}
\langle\nabla^2f\rangle_i=\sum_j 2 \frac{f_i-f_j}{r_{i j}} \boldsymbol{e}_{i j} \cdot \nabla W_{i j}V_j
\end{equation}
where $\boldsymbol{e}_{i j}=\frac{\boldsymbol{r}_{i j}}{r_{i j}}$ is a unit vector in the inter-particle direction. 
Substituting Eq. (26) into Eq. (25), the dissipation term in the SPH discrete scheme is obtained as follows,
\begin{equation}
\mathcal{D}_i=2\delta hc_s\sum_j  \frac{\rho_i-\rho_j}{r_{i j}} \boldsymbol{e}_{i j} \cdot \nabla W_{i j}V_j
\end{equation}

It should be noticed that, when utilizing Eq. (27) in a multiphase flow simulation, the density at the interface is discontinuous. This may cause significant errors in calculation of the finite-difference scheme $\frac{\rho_i-\rho_j}{r_{i j}}$ in Eq. (27), especially when the density different is large. Therefore, this dissipation term should be applied independently for each phase \cite{hammani2020detailed,sun2021accurate1,sun2021accurate2}, especially in the large density ratio cases.

\subsection{Building a generalized density dissipation term}
Based on the above analysis, we expect a more generalized density dissipation which can pass through the interface of the multiphase flow, even for the large-density-ratio ones. Besides, it should be able to recover the basic dissipation term when applied to the single phase flow. 

To successfully implement it, two changes are made to the basic dissipation term. First, according to Zheng and Chen \cite{zheng2019multiphase}, the density dissipation term in Eq. (24) can be replaced by the dissipation of the density increment, that is,
\begin{equation}
\mathcal{D}=\delta hc_s\nabla^2\widetilde{\rho}
\end{equation}
\begin{equation}
\widetilde{\rho}=\rho-\rho_0=\frac{p}{{c_s}^2}
\end{equation}
Since $c_s$ is a constant, substituting Eq. (29) into Eq. (28) yields,

\begin{equation}
\mathcal{D}=\frac{\delta h}{c_s}\nabla^2p
\end{equation}
Substituting Eq. (30) into Eq. (26), the dissipation term in the SPH discrete scheme becomes,
\begin{equation}
\mathcal{D}_i=\frac{2\delta h}{c_s}\sum_j  \frac{p_i-p_j}{r_{i j}} \boldsymbol{e}_{i j} \cdot \nabla W_{i j}V_j
\end{equation}

Compared with Eq. (27), Eq. (31) can be used to simulate multiphase flow with continuous pressure at the interface, even if the density discontinuity exists.

The second modification is the dissipation volume correction factor (VCF) which is inspired by the dissipation volume conservation. The density dissipation changes not only the particle density but also the particle volume. In the case of a particle pair comprising of particles $i$ and $j$, if the volume increment of particle $i$ is greater (smaller) than the volume decrement of particle $j$, The crowding (sparsity) distribution occurs between particle $i$ and particle $j$. It is known that particle distribution is one of the key factors to maintaining stability in SPH \cite{vacondio2021grand}. Thus to avoid crowding or sparsity particle distribution, the volume changes causing by the density dissipation in a particle pair should be balanced.

The volume change of a particle due to the density dissipation is,
\begin{equation}
\Delta V=\frac{m}{\rho^{\prime}}-\frac{m}{\rho^{\prime}+\Delta \rho}=\frac{m\Delta \rho}{\rho^{\prime}(\rho^{\prime}+\Delta \rho)}
\end{equation}
where $\rho^{\prime}$ is the fluid particle density without the correction of the density dissipation and $\Delta \rho$ is the increment of the density of the fluid particle due to the utilization of the density dissipation. Thus $\rho=\rho^{\prime}+\Delta \rho$. Considering the particle mass is aways constant, Eq. (32) finally can be rewritten as,
\begin{equation}
\Delta V=\frac{\rho_{0}V_{0}\Delta \rho}{\rho^{\prime} \rho}
\end{equation}
where $V_{0}$ are the initial density and volume of the particle. Considering $\rho_{0}\approx\rho^{\prime}$ (weakly compressible), Eq. (33) is simplified as,
\begin{equation}
\Delta V\approx\frac{\Delta \rho}{\rho}V_{0}
\end{equation}
In a time-step $\Delta t$, $\Delta \rho=\mathcal{D}\Delta t$. Thus Eq. (34) can also be represented as,
\begin{equation}
\Delta V\approx\frac{\mathcal{D}}{\rho}V_0\Delta t
\end{equation}

In the model of uniform initial particle volume $V_0$, for a particle pair consisting of particle $i$ and particle $j$, we can realize  $\Delta V_i=-\Delta V_j$ by setting $\frac{|\mathcal{D}_i|}{|\mathcal{D}_j|}=\frac{\rho_i}{\rho_j}$. To satisfy the above conditions, Eq. (31) is modified to,
\begin{equation}
\mathcal{D}_i=\frac{2\delta h}{c_s}\sum_j \frac{2\rho_i}{\rho_i+\rho_j} \frac{p_i-p_j}{r_{i j}} \boldsymbol{e}_{i j} \cdot \nabla W_{i j}V_j
\end{equation}
with the VCF $\frac{2\rho_i}{\rho_i+\rho_j}\in(0,2)$. This term tends to the maximum value of 2 or minimum value of 0 when the density ratio is large. In  section 4, we will show that this VCF greatly improves the computational stability when simulating the multiphase flow with large density ratio.

\subsection{The relationship with basic dissipation in $\delta$-SPH}

In the single phase flow simulation, all of the particles have a same initial density $\rho_0$. The following equation is satisfied due to $\rho_0$ is a constant

\begin{equation}
\nabla^2\widetilde{\rho}=\nabla^2(\widetilde{\rho}+\rho_0)=\nabla^2\rho
\end{equation}
which means that the dissipations in Eq. (27) and Eq. (31) are the same. In addition, the VCF in Eq. (36) meets $\frac{2\rho_i}{\rho_i+\rho_j}\approx1$ due to same initial densities of particles and the weakly-compressible condition. Therefore, Eq. (27) can be treated as a specific case of Eq. (36).

\subsection{The relationship with dissipation in approximate Riemann solver}
Studies have been carried out to analysis the relationship between the artificial numerical dissipation and the dissipation caused by utilizing the approximate Riemann solver in SPH. Most of them focus on the dissipation that arises in the momentum equation or the energy equation \cite{monaghan1997sph, puri2014approximate}. Recently, Green et al. \cite{green2019smoothed} recovered the dissipation of $\delta$-SPH from the continuous equation with the approximate Riemann solver for single-phase flow simulation. 

Many studies showed that SPH with the approximate Riemann solver can obtain a smooth pressure field not only in single-phase flow \cite{zhang2017weakly, xu2016improved} simulations but also in multiphase flow simulations \cite{rezavand2020weakly,meng2020multiphase, wang2021new}. Therefore, it is natural to think that the dissipation from the approximate Riemann solver scheme in the continuous equation is also generalized, which can cross the interfaces of different fluids.  

Unlike $\delta$-SPH, which directly embeds the dissipation term in the continuity equation, the dissipation induced by the Riemann approximation solver is implicit in the average velocity constructed by the numerical scheme. We give an example to extract the dissipation contribution with the structure similar to Eq. (36) from the multiphase SPH model with the approximate Riemann solver \cite{meng2020multiphase}, which would help us to clarify the relationship between the two methods.

The discrete continuous equation with the approximate Riemann solver reads  \cite{meng2020multiphase},
\begin{equation}
\frac{d \rho_i}{d t}=2 \rho_i \sum_j^N\left(\boldsymbol{v}_i-\boldsymbol{v}^*\right) \cdot \nabla_i W_{i j} V_j
\end{equation}
where $\boldsymbol{v}^*$ is the intermediate variable and satisfies
\begin{equation}
\boldsymbol{v}^*=v^* \boldsymbol{e}_{i j}+\left(\frac{\boldsymbol{v}_i+\boldsymbol{v}_j}{2}-\frac{v_L+v_R}{2} \boldsymbol{e}_{i j}\right)
\end{equation}
where $v_L=\boldsymbol{v}_j\boldsymbol{e}_{ij}$ and $v_R=\boldsymbol{v}_i\boldsymbol{e}_{ij}$. $v^*$ in Eq. (38) given by the Roe's approximate Riemann solver is
\begin{equation}
v^*=\frac{1}{2}\left[v_L+v_R+\frac{1}{C_{R L}}\left(p_L-p_R\right)\right]
\end{equation}
where $p_L=p_j$, $p_R=p_i$, and $C_{R L}$ is the Roe-averaged Lagrangian sound speed, which reads
\begin{equation}
C_{R L}=\frac{c_R \rho_R \sqrt{\rho_R}+c_L \rho_L \sqrt{\rho_L}}{\sqrt{\rho_R}+\sqrt{\rho_L}}
\end{equation}
where $\rho_L=\rho_j$ and $\rho_R=\rho_i$. Assuming $c_R=c_L=c_s$ and substituting Eqs. (39)-(41) into Eq. (38), we finally obtain 
\begin{equation}
\frac{d \rho_i}{d t}=-\rho_i \sum_j^N\left(\boldsymbol{v}_j-\boldsymbol{v}_i\right) \cdot \nabla_i W_{i j} V_j + \mathcal{D}_{Gi}
\end{equation}
where $\mathcal{D}_{Gi}$ is the dissipation term originating from the implementation of the Roe's approximate Riemann solver, which reads,
\begin{equation}
\mathcal{D}_{Gi}=\frac{2}{c_s}\sum_j \frac{\rho_i(\sqrt{\rho_i}+\sqrt{\rho_j})}{\rho_i\sqrt{\rho_i}+\rho_j\sqrt{\rho_j}} (p_i-p_j)\boldsymbol{e}_{i j} \cdot \nabla W_{i j}V_j
\end{equation}
By multiplying $\frac{r_{ij}} {r_{ij}}$ in the sum operator in Eq. (43), we have
\begin{equation}
\mathcal{D}_{Gi}=\frac{2}{c_s}\sum_j r_{ij} \frac{\rho_i(\sqrt{\rho_i}+\sqrt{\rho_j})}{\rho_i\sqrt{\rho_i}+\rho_j\sqrt{\rho_j}} \frac{p_i-p_j}{r_{i j}}\boldsymbol{e}_{i j} \cdot \nabla W_{i j}V_j
\end{equation}
Now we go back to the present generalized density dissipation $\mathcal{D}_i$ in Eq. (36). Moving $h$ into the sum operator in Eq. (36) yields
\begin{equation}
\mathcal{D}_i=\frac{2\delta}{c_s}\sum_j h\frac{2\rho_i}{\rho_i+\rho_j} \frac{p_i-p_j}{r_{i j}} \boldsymbol{e}_{i j} \cdot \nabla W_{i j}V_j
\end{equation}
By Comparing Eq. (44) with Eq. (45), we can found that both equations have a Laplacian of the pressure $p$, referring the discrete structure $\nabla^2p=\sum_j  \frac{p_i-p_j}{r_{i j}} \boldsymbol{e}_{i j} \cdot \nabla W_{i j}V_j$. Moreover, the first terms in the sum operators in Eqs. (44) and (45), $r_{ij}$ and $h$, are in the same order. Note that $\mathcal{D}_{Gi}$ cannot be controlled artificially. However, one can vary $\delta$ in $\mathcal{D}_{i}$ to adjust the amount of dissipation. From the point of view of dissipation volume conservation, the generalized dissipation term proposed in this paper combines the VCF $\frac{2\rho_i}{\rho_i+\rho_j}$. The dissipation of the SPH model with the Roe's approximate Riemann solver also includes a correction factor $\frac{\rho_i(\sqrt{\rho_i}+\sqrt{\rho_j})}{\rho_i\sqrt{\rho_i}+\rho_j\sqrt{\rho_j}}$, which originates from the Roe-averaged Lagrangian sound speed. This correction factor varies when applying different approximate Riemann solvers, e.g., Rusanov flux \cite{xu2016improved}, primitive variable Riemann solver \cite{fang2022accurate}.

From the above analysis, it can be seen that the numerical dissipation generated by using the approximate Riemann solver and the generalized density dissipation proposed in this paper are essentially the same kind of dissipation. Both of them arise from the process to form a function of $\nabla^2p$ to eliminate the oscillation of $p$.

\subsection{The final generalized density dissipation term discretized by SPH}
Although Eqs. (30), (31), and (36) are employed to analyze the function of the present generalized dissipation and the relationship with other numerical schemes, in practical application we discretize Eq. (28) to obtain a density dissipation term. Using the dissipation volume modification term and the SPH discrete method proposed by Antuono \cite{antuono2010free} to retain high-order terms, the generalized density dissipation term in Eq. (28) is discretized as,
\begin{equation}
\mathcal{D}_i=-2 \delta h c_s \sum_j\frac{2 \rho_i}{\rho_i+\rho_j}\left[\left(\widetilde{\rho}_j-\widetilde{\rho}_i\right)-\frac{1}{2}\left(\langle\widetilde{\rho}\rangle_j^L-\langle\widetilde{\rho}\rangle_i^L\right) \cdot \boldsymbol{r}_{i j}\right]  \frac{\boldsymbol{r}_{i j} \cdot \nabla_i W_{i j}}{\left|\boldsymbol{r}_{i j}\right|} V_j
\end{equation}
\begin{equation}
\langle\widetilde{\rho}\rangle_i^L=\sum_j\left(\widetilde{\rho}_j-\widetilde{\rho}_i\right) \boldsymbol{L}_i \nabla_i W_{i j} V_j
\end{equation}
\begin{equation}
\boldsymbol{L}_i=\left[\sum_j\left(\boldsymbol{r}_j-\boldsymbol{r}_i\right) \otimes \nabla_i W_{i j} V_j\right]^{-1}
\end{equation}

Within the scope of the kernel function, the central particle is most affected by the nearest particle (ideally, the distance between the central particle and the nearest particle is around $dx_0$, where $dx_0 \times dx_0 \times dx_0$ is the initial size of the particle), and $h$ is of the same order as $dx_0$ (we set $h=1.1dx_0$ for all the cases in this paper), so Eq. (46) can be simplified to,

\begin{equation}
\mathcal{D}_i=-\delta c_s \sum_j\frac{2 \rho_i}{\rho_i+\rho_j}\left[\left(\widetilde{\rho}_j-\widetilde{\rho}_i\right)-\frac{1}{2}\left(\langle\widetilde{\rho}\rangle_j^L-\langle\widetilde{\rho}\rangle_i^L\right) \cdot \boldsymbol{r}_{i j}\right] \boldsymbol{r}_{i j} \cdot \nabla_i W_{i j}  V_j
\end{equation}
Eq. (49) is the final discrete generalized density dissipation term which is added to the right hand of continuous equation. The ability of this dissipation to stabilize the solutions in both the multiphase and single-phase simulations will be shown in the numerical examples in the next section.

\section{Numerical examples}

Numerical examples are simulated by the present method for evaluation of its accuracy and stability in this section. In order to compare with experimental results or other simulated results, the wave elevation monitoring algorithm is mainly required in the current work. This calls for free-surface or interface capture algorithms to detect the surface particle. In the present study, we use the algorithm proposed by Doring \cite{doring2005developpement} for the free-surface capture and the algebraic indicator method proposed by Zheng et al. \cite{zheng2021novel} for the interface capture. Additionally, the artificial coefficient $\delta$ always takes 0.5 in all the examples unless otherwise specified.

\subsection{Sloshing under external excitations}

Liquid sloshing in a cuboid tank under the external excitation is considered first to test the convergence and the accuracy of the present SPH model. 

\begin{figure}[h]
\begin{center}
\includegraphics[width=1\textwidth]{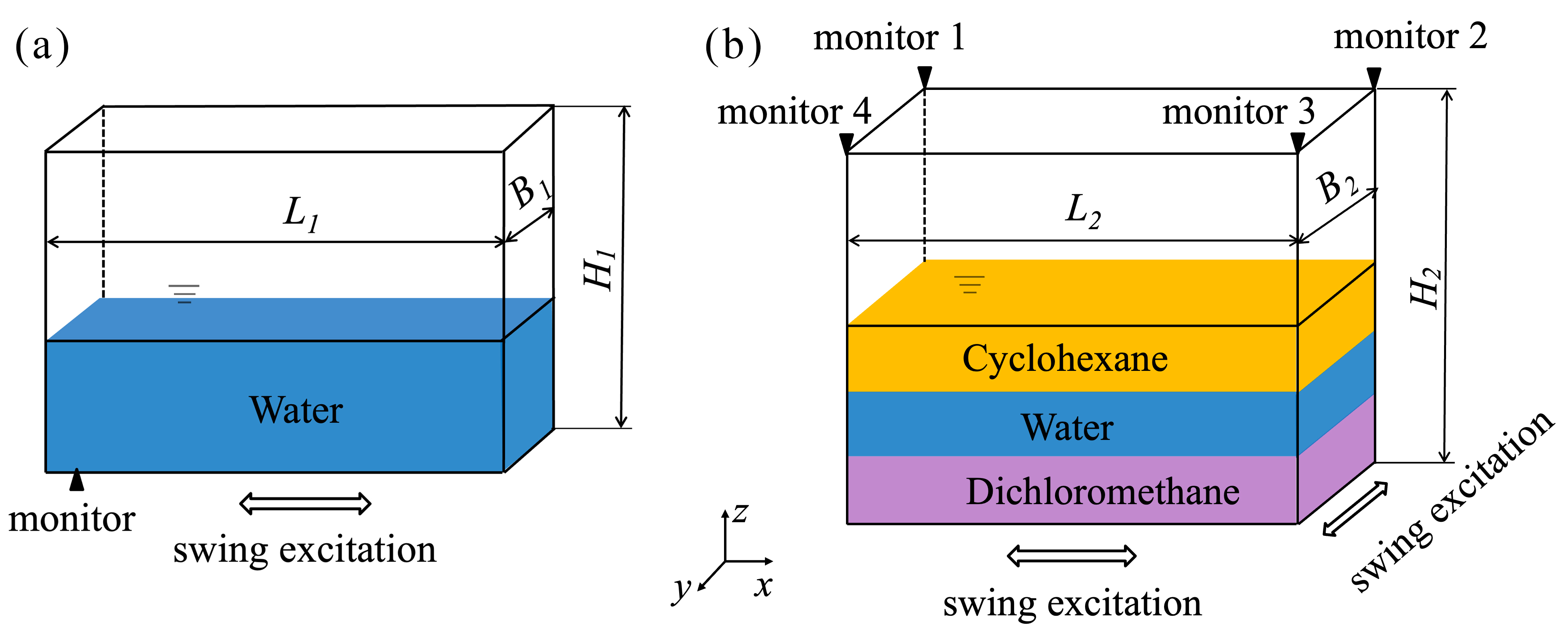}
\caption{Initial setup of liquid sloshing in a cuboid tank: (a) the single-phase case and (b) the multiphase case.} \label{fig1}
\end{center}
\end{figure}

\subsubsection{Single-phase sloshing under one-direction external excitation}

We consider a single-phase sloshing under one-direction external excitation which is experimentally studied by Lin and Liu \cite{liu2009three}. The configuration of this example is illustrated in Fig. 1 (a). A partially filled cuboid tank is subject to the periodic swing excitation. Specifically, the length ($L_1$),  height ($H_1$) and breadth ($B_1$) of the tank are $0.57m$, $0.3m$, and $0.06m$, respectively. The tank is initially filled with $0.15m$ deep of water ($h_w$) whose density is $1000kg/m^3$ and dynamics viscosity is $1.0\times10^{-3}Pa\cdot s$. The origin of the coordinate system is set at the midpoint of the intersection line between the bottom surface of the tank and the left side surface.
The swing motion of the tank follows
\begin{equation}
x=A \sin \omega t
\end{equation}
where the amplitude $A$ is chosen as $0.005m$, and the frequency $\omega$ is set to be $6.0578 s^{-1}$, which equals the nature frequency of this sloshing system. In this simulation, a monitor that tracks the elevation of free surface waves is placed $0.02m$ from the origin of the coordinate system.

\begin{figure}[h]
\begin{center}
\includegraphics[width=1\textwidth]{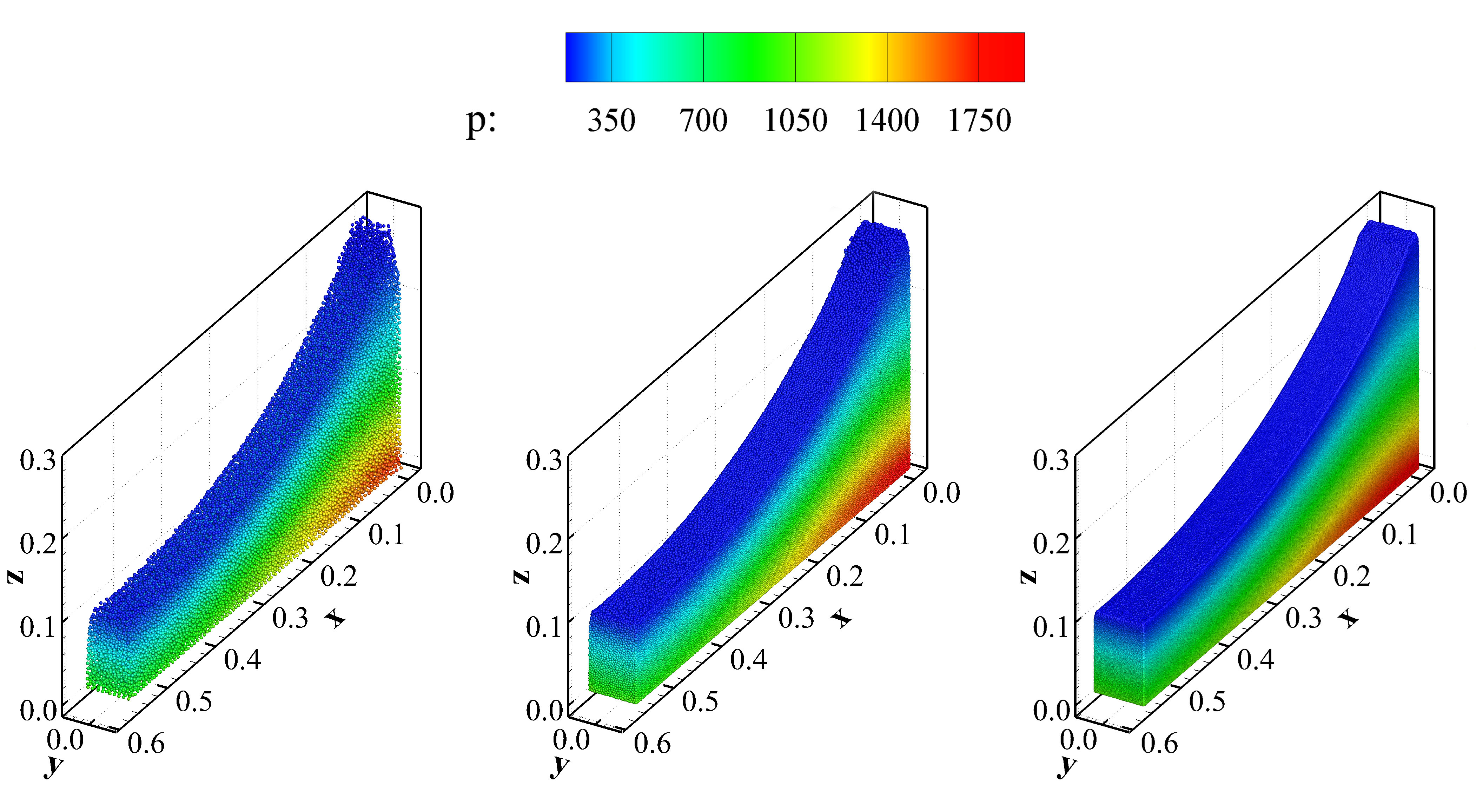}
\caption{Pressure fields for the single-phase sloshing obtained by the present SPH method with the particle sizes of $h_w/dx_0=30, 60, 120$ from the left to the right at $t=6.8s$.}\label{fig2}
\end{center}
\end{figure}

Three particle sizes, namely, $dx_0=0.005m$ ($h_w/dx_0=30$), $dx_0=0.0025m$ ($h_w/dx_0=60$) and $dx_0=0.00125m$ ($h_w/dx_0=120$) are chosen to validate the convergence of the present method. Figure 2 illustrate the pressure field obtained by the present model with the above three particle sizes. The pressure distribution is smooth for all three particle sizes and the resolution of the pressure distribution increases with the discretization precision. This shows that the generalized dissipation term can well suppress the pressure oscillation in single-phase flow simulation. This is expected as analyzed in Section 3.3, because the present dissipation term can be simplified into the dissipation term in $\delta$-SPH when simulating single-phase flow. 

\begin{figure}[H]
\begin{center}
\includegraphics[width=0.9\textwidth]{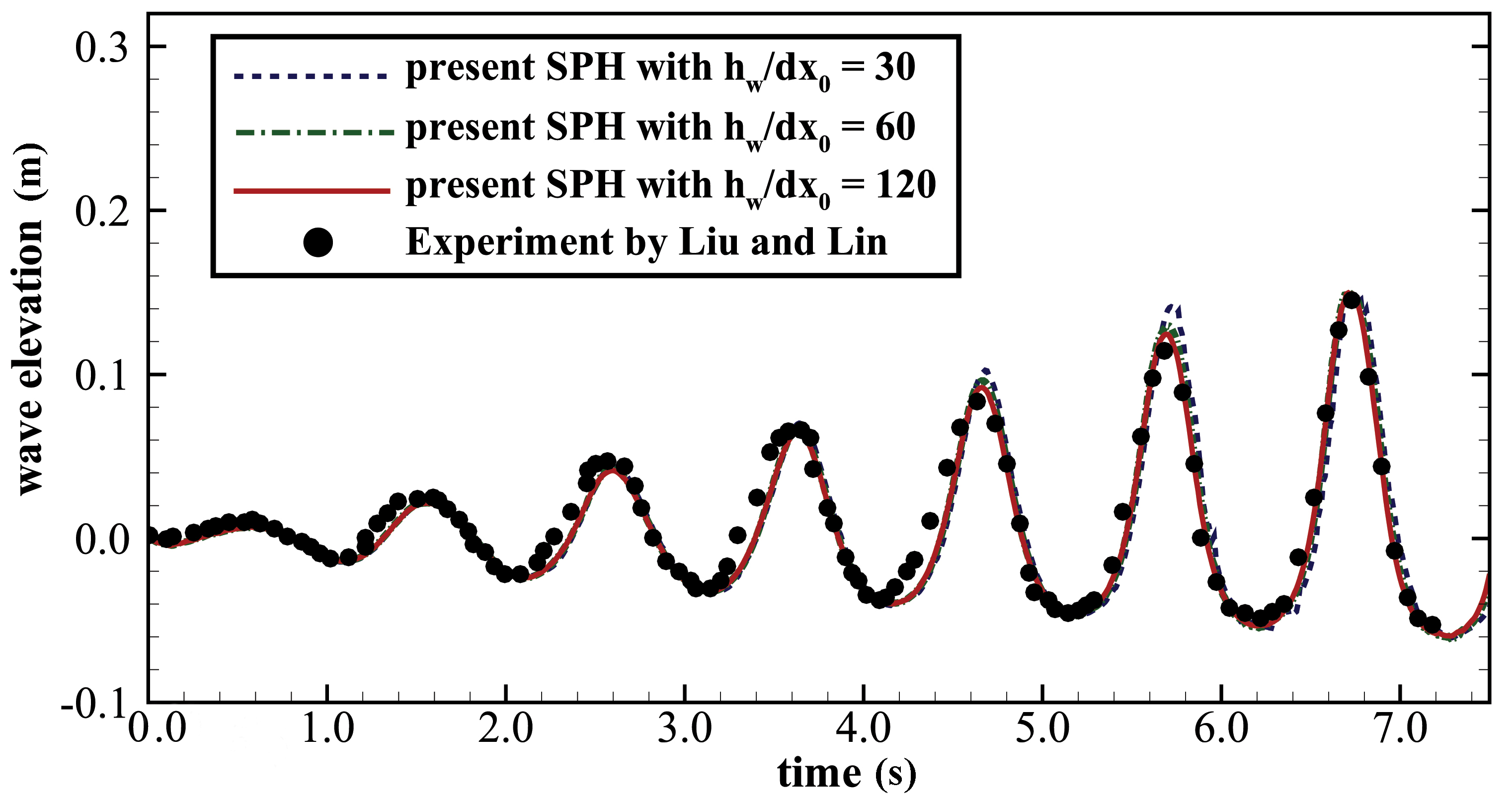}
\caption{Time history of wave elevation measured at 0.02m from the left wall for the single-phase sloshing. The present results: blue dash line for $h_w/dx_0=30$, green dash-dot line for $h_w/dx_0=60$, and red solid line for $h_w/dx_0=120$; the experimental data of Liu and Lin \cite{liu2009three}: dark dots.}\label{fig3}
\end{center}
\end{figure}

The time history of the wave elevation is recorded in Fig. 3 and compared with the the experimental data reported by Liu and Lin \citep{liu2009three}. It can be seen that all the results simulated by three particle sizes show very good agreements with the experimental data and with the increase of the particle resolution, the numerical results are more consistent with the experimental results, which validates the convergence and accuracy of the present method.

\subsubsection{multiphase sloshing under two-direction external excitations}
Although the example in Section $4.1.1$ is simulated in a 3D computational domain, it shows 2D phenomenon. A truly 3D problem, multiphase sloshing under two-direction external excitations is considered in this subsection. As illustrated in Fig. 1 (b), a cuboid tank with the length $L_2 = 1.5m$, the breadth $B_2 = 1m$, and the height $H_2 = 2m$ is filled with three layers of different liquids namely dichloromethane, water, and cyclohexane, from the bottom to the top. The height of each layer is $20cm$. The properties of the liquids are listed in Table 1, which follow those used in the experiment of Molin et al. \citep{molin2012experimental}. The particle size adopting in this case  is $0.01m$. Four wave elevation monitors are located at the corners of the four sides of the cuboid tank.  The swing motions of the tank in the $x$ and $y$ directions are

\begin{equation}
x=-B \cos \omega_1 t+B
\end{equation}
\begin{equation}
y=-B \cos \omega_2 t+B
\end{equation} 
where the amplitude $B$ is $0.0425m$, and $\omega_1 = 4.18s^{-1}$ and $\omega_2 = 4.14s^{-1}$ are the frequencies in the $x$ and $y$ directions, respectively.

\begin{table}[h]
\centering
\caption{Properties of fluids}\label{tab1}
\setlength{\tabcolsep}{8mm}
\begin{tabular}{ccc} 
\toprule
                & density & dynamic viscosity\\
\midrule
Cyclohexane     & $780~kg/m^3$ & $1.014 \times 10^{-3}~Pa \cdot s$ \\
Water           & $1000~kg/m^3$ & $1.0 \times 10^{-3}~Pa \cdot s$ \\
Dichloromethane & $1300~kg/m^3$ & $3.9 \times 10^{-4}~Pa \cdot s$ \\
\bottomrule
\end{tabular}
\end{table}

Figure 4 (a) shows the pressure distributions obtained by the present method at $t=1.81s$ and $t=2.83s$. The smooth pressure fields demonstrate the ability of the present generalized density dissipation to suppress the pressure oscillation. Figures 4 (b) and (c) are phase distributions obtained by the present SPH model and STAR-CCM+ based on the finite volume method(FVM) \cite{eymard2000finite} with the volume of fluid (VOF) \cite{hirt1981volume} for surface tracking. In such a flow problem with large deformed free surface, the results obtained the two methods are very consistent. Quantitative comparisons for wave elevation obtained by the present SPH method (lines) and STAR-CCM+ (points) are shown in Fig. 5. The time histories of the wave elevation of the free-surface and the cyclohexane-water interface recorded by four monitors are presented in Figs. 5 (a) and (b), respectively. The good agreement between the results obtained by the two methods proves the accuracy of the present SPH model for simulating the multiphase flow with free-surface.

\begin{figure}[H]
\begin{center}
\includegraphics[width=0.88\textwidth]{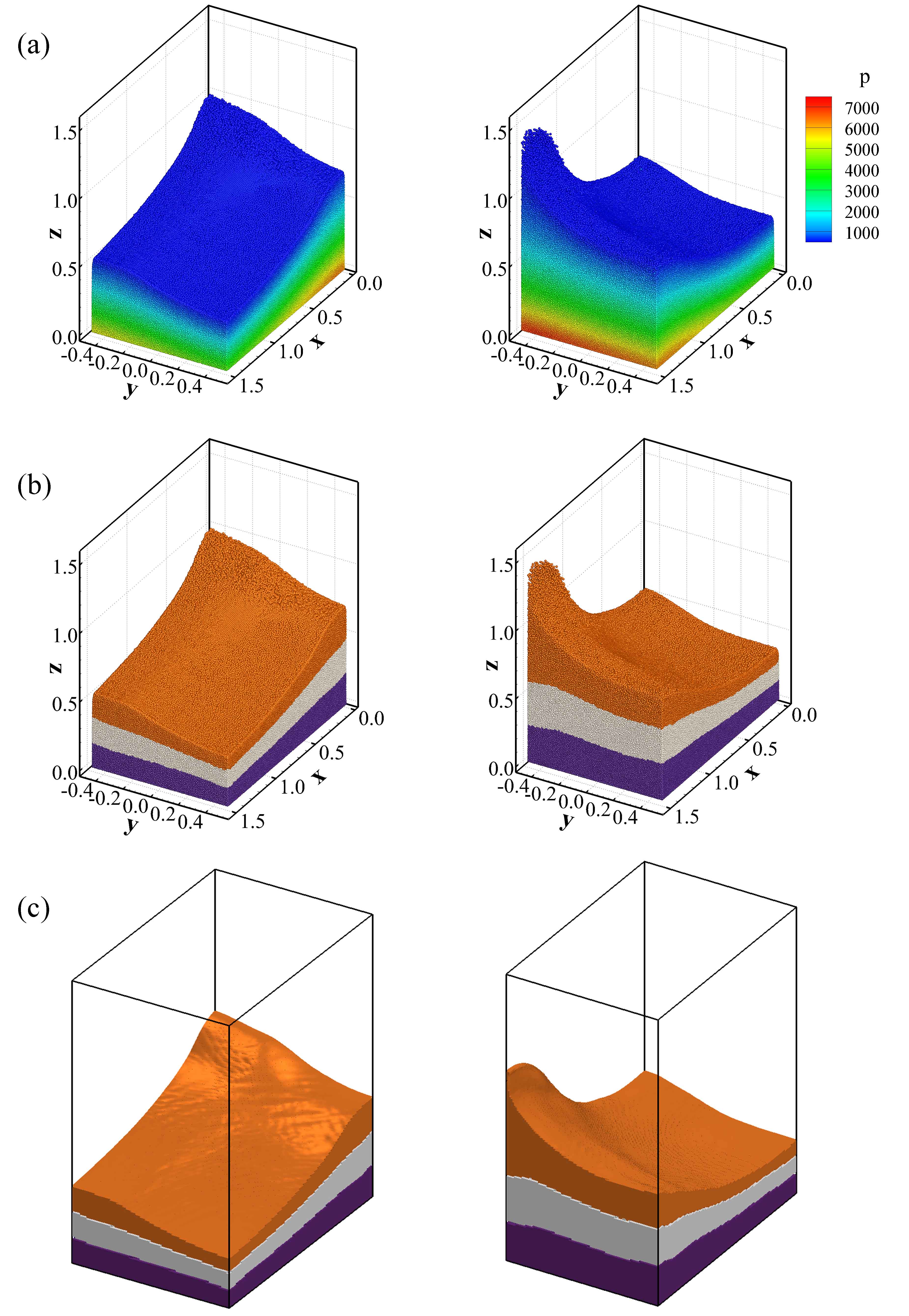}
\caption{Numerical results of the multiphase sloshing at $t=1.81s$ (left) and  $t=2.83s$ (right): (a) pressure field and (b) phase distribution obtained by the present SPH method, and (c) phase distribution obtained by STAR-CCM+.}\label{fig4}
\end{center}
\end{figure}

\begin{figure}[H]
\begin{center}
\includegraphics[width=1\textwidth]{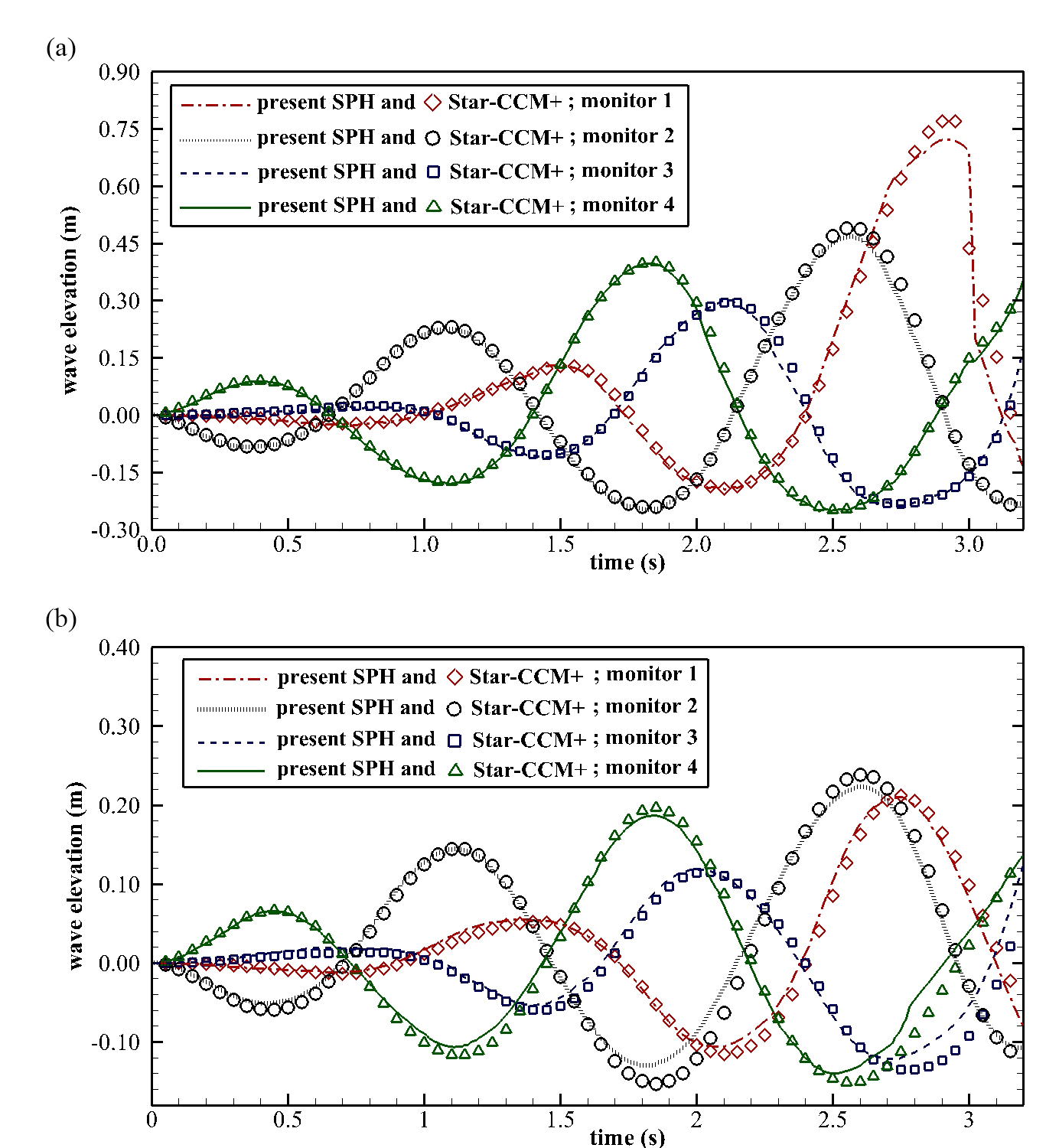}
\caption{Time history of wave elevation for the  multiphase sloshing: (a) free-surface of cyclohexane and (b) interface of cyclohexane and water. Lines represent the simulation results obtained by the present SPH method while points represent the simulation results obtained by STRA-CCM+. Different colors represent the data recorded by different monitors as shown in Fig. 1.}\label{fig5}
\end{center}
\end{figure}

\subsection{Bubble rising}

The benchmarks of single and double bubbles rising are simulated to test the performance of the generalized dissipation term for multiphase flow with large density ratio. These two cases has been experimentally studied by Brereton and Korotney \cite{brereton1991coaxial} and numerically investigated by both the mesh methods \cite{sitompul2019filtered, van2005numerical, yu2016improved} and the particle methods \cite{zhang2015sph, yan2020higher}. 

\begin{figure}[H]
\begin{center}
\includegraphics[width=1\textwidth]{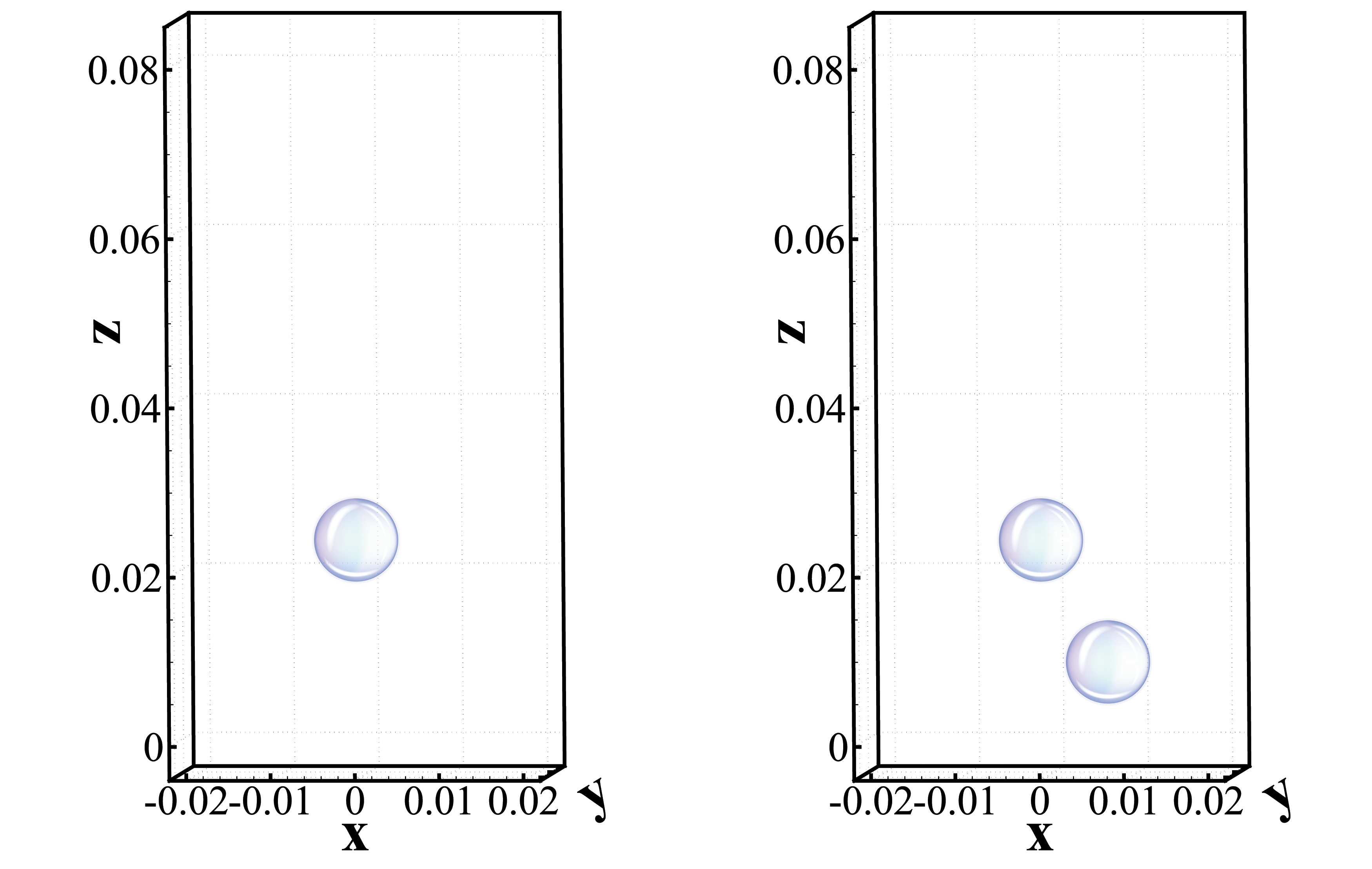}
\caption{Initial setup for the cases of single bubble rising (left) and double bubbles rising (right).}\label{fig6}
\end{center}
\end{figure}

It is worth noting that the VCF, $\frac{2\rho_i}{\rho_i+\rho_j}$ is no longer approximately equal to 1 at the interface due to the large density ratio. Thus, in this subsection, we show the improvement of the stability by incorporating the VCF in the generalized density dissipation. 

\begin{table}[H]
\centering
\caption{Parameters for the cases of the single and double bubbles rising}\label{tab2}
\setlength{\tabcolsep}{8mm}
\begin{tabular}{cc} 
\toprule
Computational domain        & $0.02~m~\times~0.02~m~\times~0.04~m$  \\
Bubble radius $R$           & $0.005~m$  \\
Initial bubble position     &  \\
single bubble               & $(x_0, ~y_0, ~z_0)=(0~m,~0~m,~0.025~m)$ \\
double bubbles              & $\left\{\begin{array}{l}\left(x_1,~ y_1, ~z_1\right)=(0~m,~0~m,~0.025~m) \\ \left(x_2,~ y_2,~ z_2\right)=(0.008~m,~ 0~m,~0.01~m)\end{array}\right.$  \\
Liquid density   $\rho_l$   & $1000~kg/m^3$  \\
Bubble density   $\rho_b$   & $10~kg/m^3$  \\

Liquid viscosity $\mu_l$    & $4.63~\times~10^{-2}~Pa \cdot s$\\
Bubble viscosity $\mu_b$    & $4.63~\times~10^{-4}~Pa \cdot s$\\
Surface tension  $\sigma$   & $0.0606~N/m$\\
\bottomrule
\end{tabular}
\end{table}

Both cases are simulated in the domain of $[0.02m \times 0.02m \times 0.04m]$ and all the bubbles have the same radius $R=0.005m$, as shown in Fig. 6. The parameters used in these cases are listed in Table 2. The control parameters, namely, the Morton and Eotvos numbers, are defined as:
\begin{equation}
Mo=\frac{g \mu_l^4 (\rho_l-\rho_b)}{\rho_l^2 \sigma^3}=2 \times 10^{-4}
\end{equation}

\begin{equation}
Eo=\frac{4(\rho_l-\rho_b) g R^2}{\sigma}=16
\end{equation}
where $g$ is the gravity; $\rho_l$ and $\rho_b$ are the density of the liquid and the bubble, respectively.

To prevent unphysical infiltration of particles in different phases, a repulsive force at the interface is widely used in the SPH simulation of bubble rising problem \cite{colagrossi2003numerical,grenier2009hamiltonian,zhang2015sph,zheng2019multiphase}. In the present study, this forcing term is also added to the right hand of the momentum equation, which reads,
\begin{equation}
\boldsymbol{F}_i^{\text {repulsive }}=-\beta _i^j \sum_j\left(\frac{p_i}{\Gamma_i}+\frac{p_j}{\Gamma_j}\right) \nabla_i W_{ij} V_j
\end{equation}
\begin{equation}
\beta _i^j= \begin{cases}0.08 & \text { if particles } i \text { and } j \text { belong to different phases } \\ 0 & \text { if particles } i \text { and } j \text { belong to the same phase. }\end{cases}
\end{equation}

The bubble rising velocity $U_{bubble}$ is calculated by the following formulation:
\begin{equation}
U_{bubble}= \frac{\sum z_{t+\Delta t} -\sum z_{t} }{N_b\Delta t}
\end{equation}
where $N_b$ represents the total number of bubble particles and  $z_{t+\Delta t}$ and $z_{t}$ represent the z-direction coordinates of a bubble particle at ${t+\Delta t}$ and $t$, respectively.

\subsubsection{Single bubble rising}

Three sizes of particles, i.e., $dx_0=0.001m$ ($R/dx_0=5$), $dx_0=0.0005m$ ($R/dx_0=10$), and $dx_0=0.00025m$ ($R/dx_0=20$) are utilized to test the convergence of the present method. Figure 7 shows the phase distributions obtained by above particle sizes at $t=0.005s$, $0.09s$, and $0.16s$. The calculation results show very good consistency. Therefore, we choose the middle particle size of $dx_0=0.0005m$ ($R/dx_0=10$) for subsequent simulations.

\begin{figure}[H]
\begin{center}
\includegraphics[width=1\textwidth]{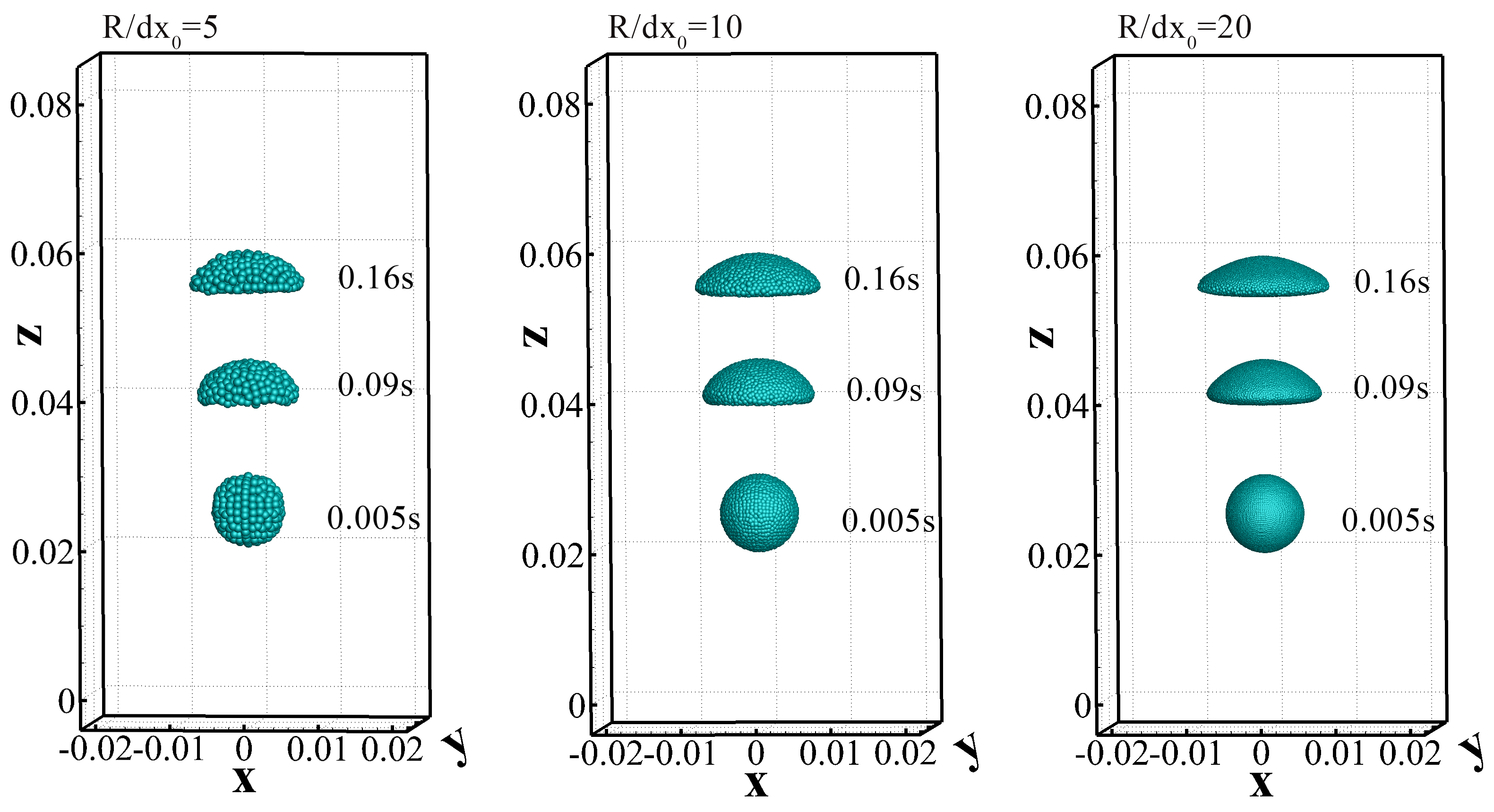}
\caption{Phase distributions at $t=0.005s$, $0.09s$, and $0.16s$ obtained by the present SPH model with the particle sizes of $R/dx_0=$5, 10, 20 from the left to the right.}\label{fig7}
\end{center}
\end{figure}

The effect of the VCF on numerical stability when simulating large-density-ratio multiphase flow is tested. The pressures on the bubble surface as well as the  boundaries of the computational domain at $t=1.6s$ are illustrated in Fig. 8. The results in Fig. 8 (a) are obtained by the SPH model utilizing a generalized dissipation term combined with the VCF. For $\delta$ ranging from $0.25$ to $3$, the contours of the bubble surface pressure and the boundary pressure are very smooth and the bubble shapes are consistent. The effects of $\delta$ on the pressure distribution and the bubble shape are negligible. Figure 8 (b) shows the results obtained by the SPH model utilizing the generalized dissipation term without the VCF. For $\delta$ ranging from $0.25$ to $3$, although the boundary pressure distributions are very smooth and the effect of $\delta$ on the boundary pressure distribution is very small, the shape of the bubble surface and the pressure distribution on the bubble surface does show some unexpected variations. The effects of $\delta$ on the shape and the pressure distribution of the bubble surface is quite noticeable, which indicate strong numerical instability and the interface. Figure 9 shows the time history of the bubble rising velocity obtained by using the present dissipation term with and without the VCF under different $\delta$ (denoted by different lines), as well as that obtained by Zhang et al. \cite{zhang2015sph} (denoted by  dots). With the VCF in the dissipation term, when $\delta$ varies from 0.125 to 3, the rising velocities of the bubbles varies smoothly and approach constant value, which are very consistent with the numerical result of Zhang et al. \cite{zhang2015sph}. However, without the VCF in the dissipation term,  even within a narrower variation of $\delta$  (from 0.0625 to 1), the curves always show oscillation and are hard to consistent

This can be interpreted as, when the VCF based on the dissipation volume conservation is applied, the total volume of the particle pair will not be affected by the artificial dissipation, which maintains a uniform particle distribution in the support domain and, therefore, greatly improves the numerical stability of the present method.

\begin{figure}[H]
\begin{center}
\includegraphics[width=1\textwidth]{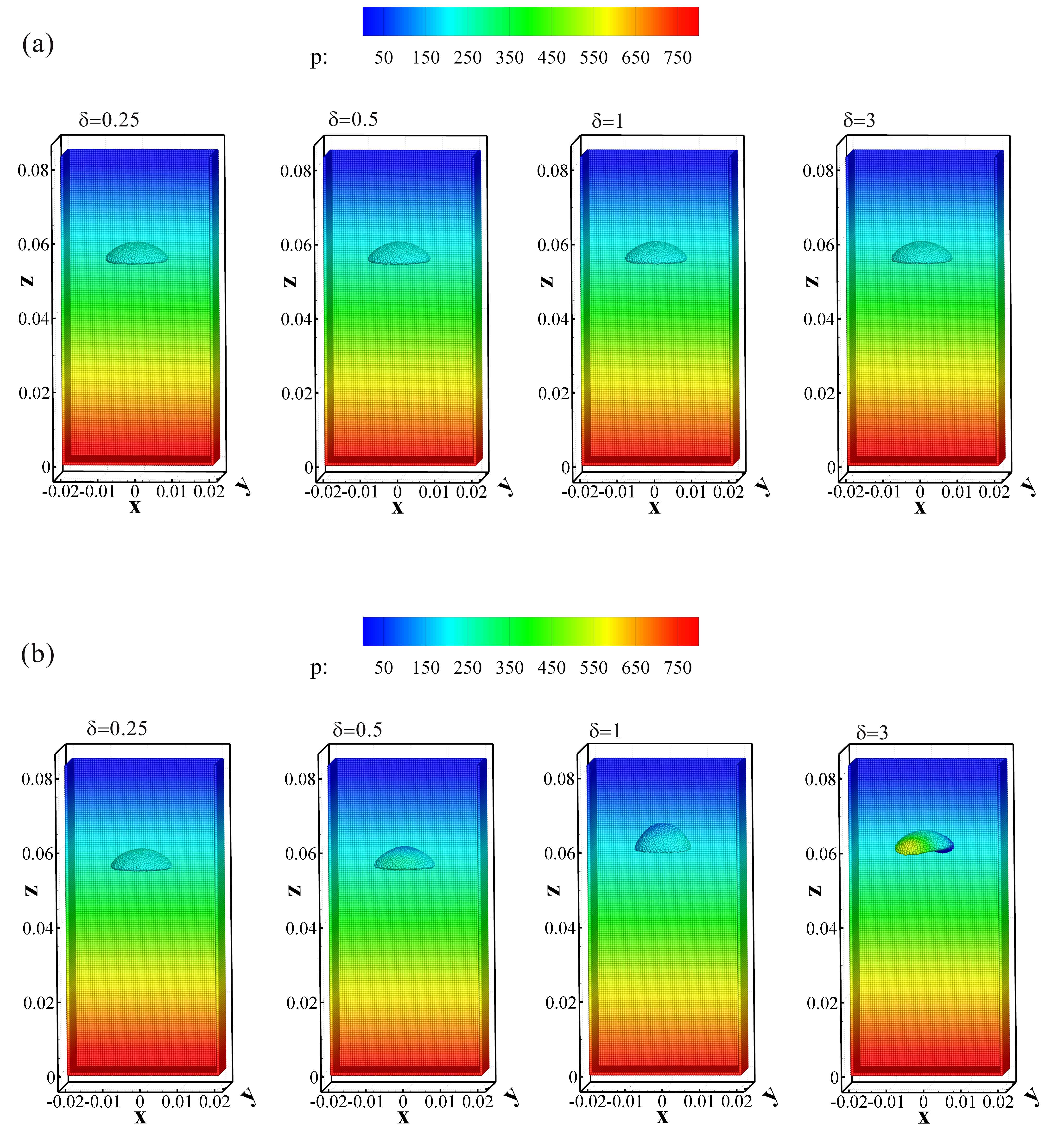}
\caption{Pressure fields for $\delta=$ 0.25, 0.5, 1, 3 (from the left to the right) at $t=0.16s$ by utilizing the dissipation terms (a) with the volume correction factor(VCF) and (b) without the volume correction factor(VCF).}\label{fig8}
\end{center}
\end{figure}

\begin{figure}[H]
\begin{center}
\includegraphics[width=0.9\textwidth]{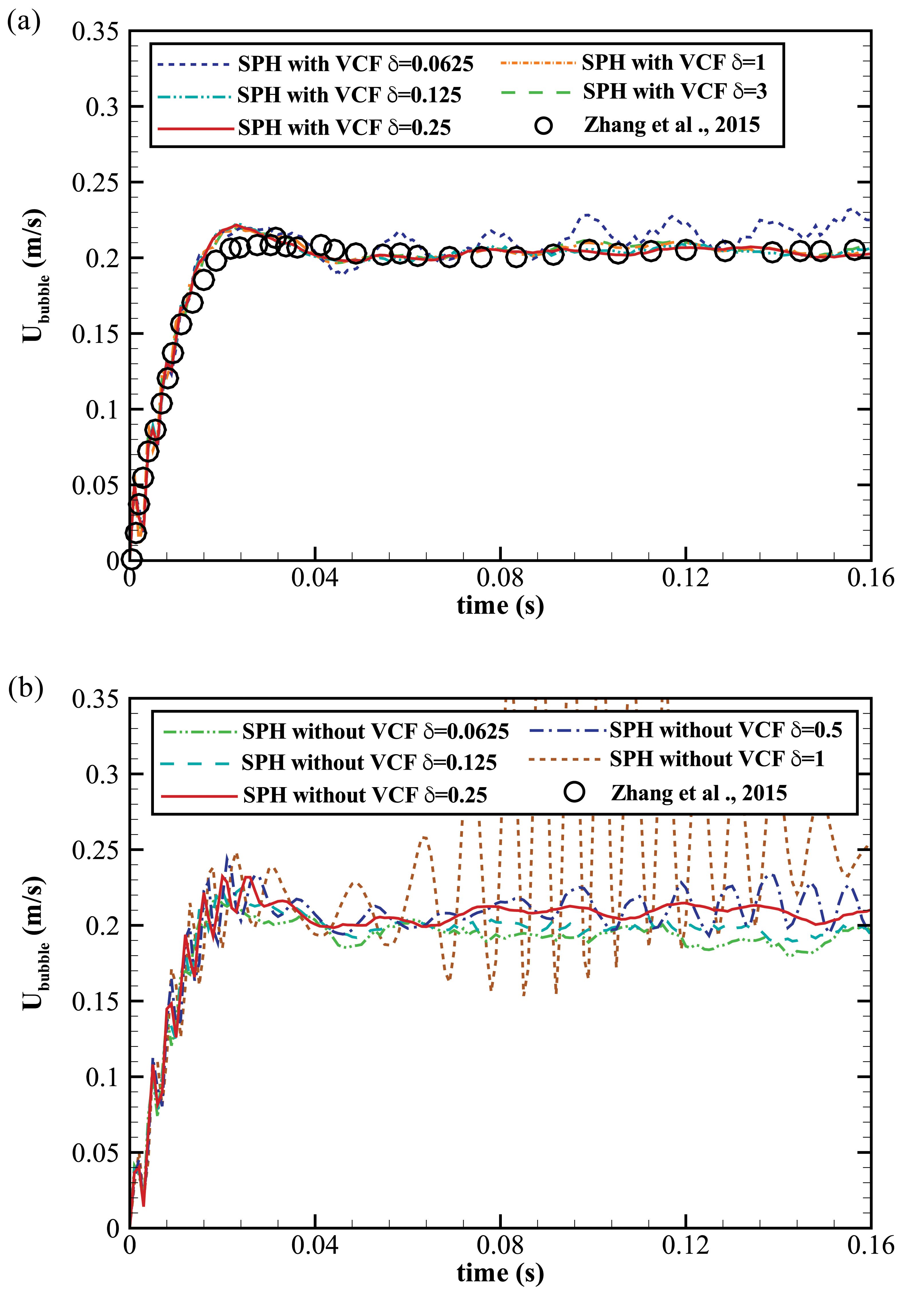}
\caption{Time histories of the bubble rising velocity obtained by Zhang et al. \cite{zhang2015sph} (denoted by dots) and the generalized density dissipation terms (a) with the volume correction factor(VCF) and (b) without the volume correction factor(VCF) (denoted by lines).}\label{fig9}
\end{center}
\end{figure}

\subsubsection{Double bubbles rising}
Double bubbles rising includes the processes of bubble chasing, contacting, and coalescence so that their interfaces suffer from more complex morphology and pressure changes than those for the single bubble rising. We use this example to verify the robustness of the present generalized dissipation term. 

\begin{figure}[H]
\begin{center}
\includegraphics[width=1\textwidth]{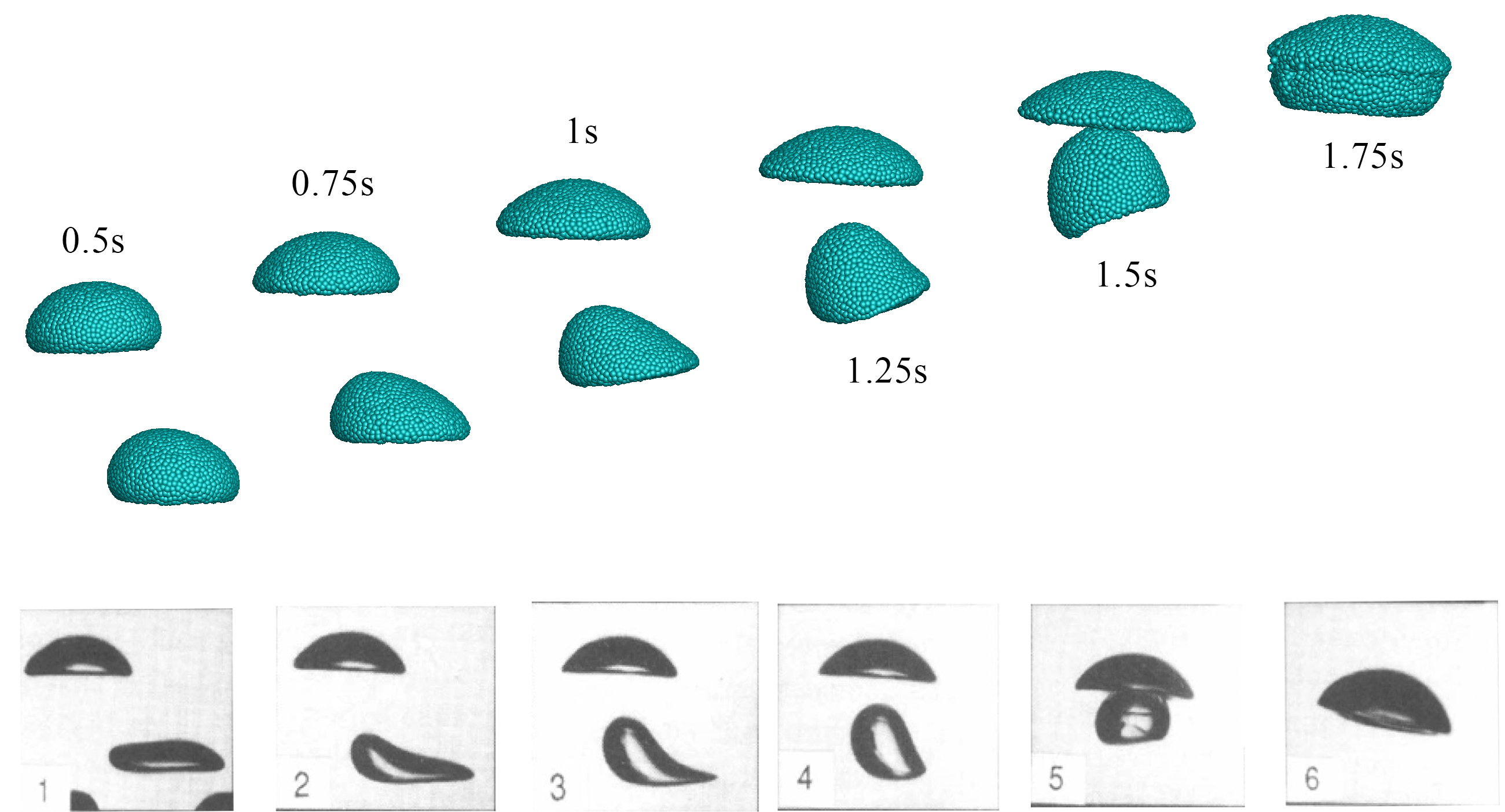}
\caption{Snapshots at different time instants during double bubbles rising; top: the present results; bottom: the results of Brereton and Korotney \cite{brereton1991coaxial}.}\label{fig10}
\end{center}
\end{figure}

Figure 10 shows the bubble shape and position obtained by the present SPH model at different time instants and the corresponding experimental snapshots obtained by Brereton and Korotney \cite{brereton1991coaxial}. The typical features of the processes of double bubbles rising, i.e., chasing, the tail bubble catching up with the leading bubble, and coalescence are clearly captured by the present method, which are consistent with the experimental snapshots . The comparison of the average rising velocity of bubbles obtained by the present SPH model, the unfiltered and filtered cumulant LBMs \cite{sitompul2019filtered} is shown in Fig. 11, where the good agreement can be observed. 

\begin{figure}[H]
\begin{center}
\includegraphics[width=1\textwidth]{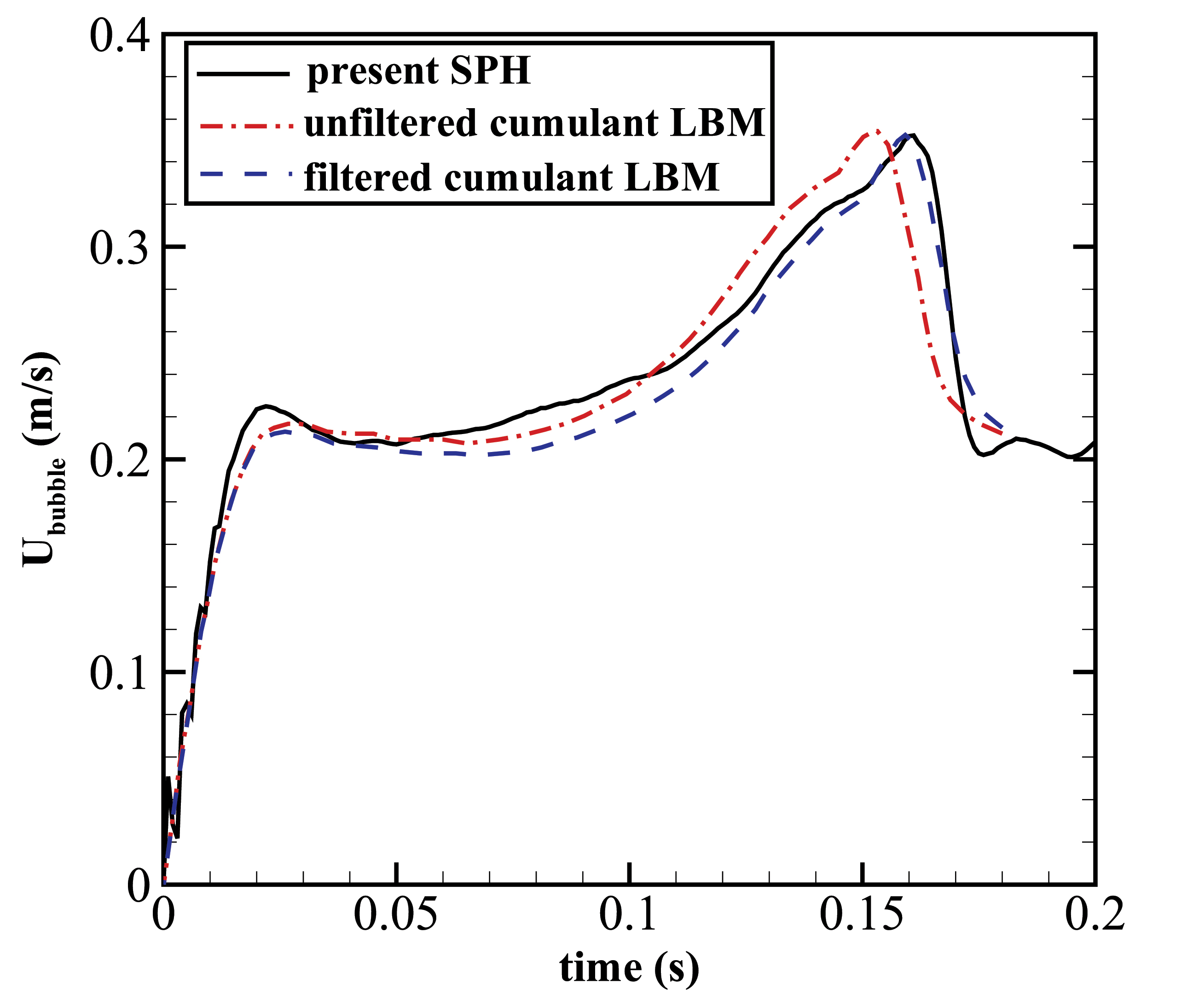}
\caption{Time history of the average rising velocity of the two bubbles obtained by the present SPH model and the unfiltered and filtered cumulant LBMs of Sitompul and Aoki \cite{sitompul2019filtered}.}\label{fig11}
\end{center}
\end{figure}

Figure 12 shows the predicted pressure fields for four typical time instants during the double bubbles rising, including initial rising ($t=0.005s$), chasing ($t=0.1s$), touching ($t=0.15s$), and coalescence ($t=0.175s$). The pressure distributions on the bubble surface and the domain boundaries are excellently smooth, which demonstrate the capability of the present generalized density dissipation term on suppressing the pressure oscillation.

\begin{figure}
\begin{center}
\includegraphics[width=1\textwidth]{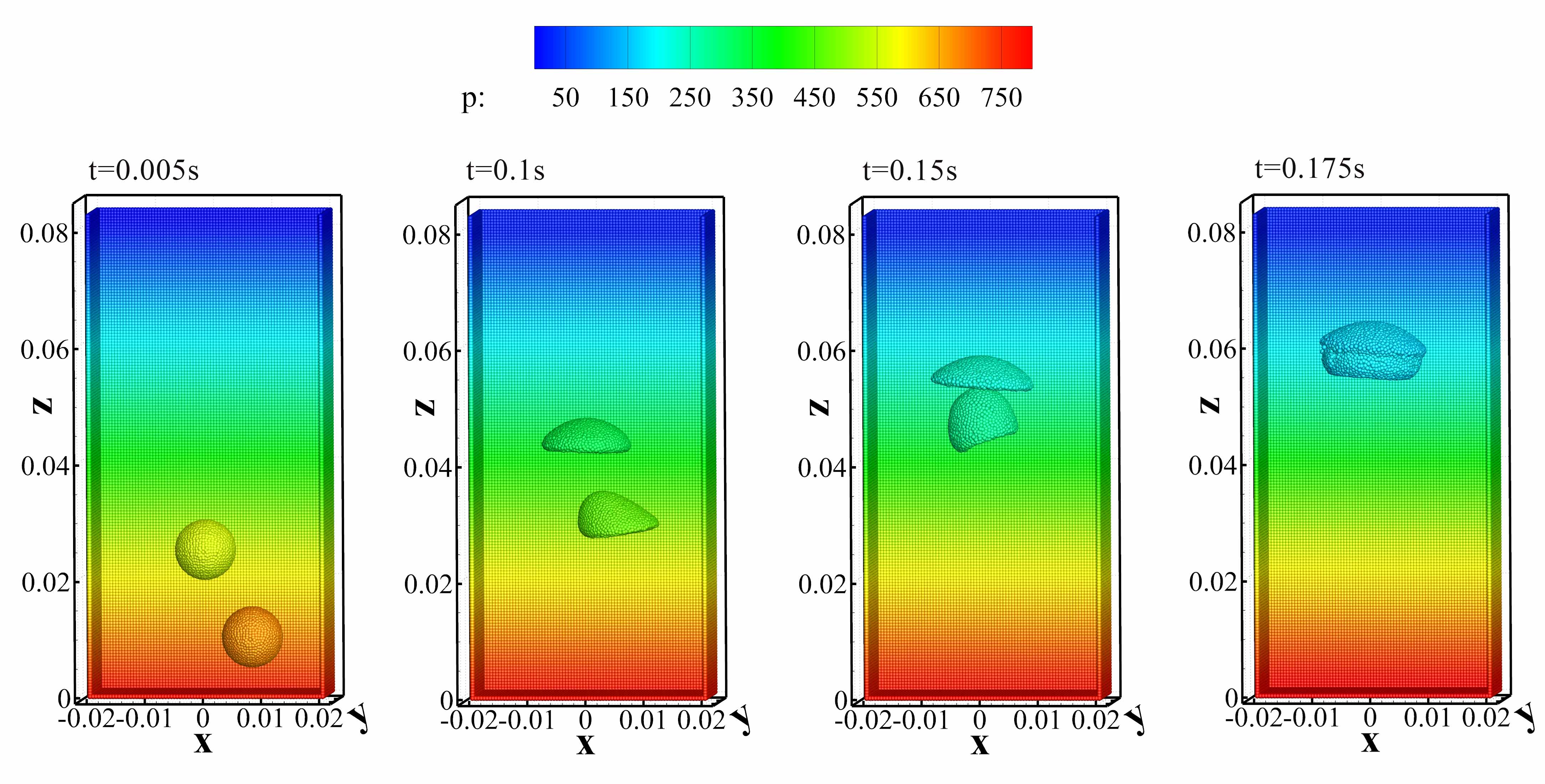}
\caption{Pressure fields obtained by the present SPH model for four typical time instants during the double bubbles rising, including initial rising ($t=0.005s$), chasing ($t=0.1s$), touching ($t=0.15s$), and coalescence ($t=0.175s$).}\label{fig12}
\end{center}
\end{figure}

\subsection{Rayleigh-Taylor instability}

The classic multiphase flow benchmark, the 3D Rayleigh-Taylor instability,  is simulated in this section. The problem setup in this paper follows that reported in Refs. \cite{he1999three,lee2013numerical,hamzehloo2021direct}. The computational domain $\Omega=x \times y \times z=(0,1m) \times(0,1m) \times(0,4m)$ is initially divided by the interface at $z_0$ defined as,
\begin{equation}
z_0=2 +0.05\left[\cos 2 \pi x+\cos 2 \pi y\right]
\end{equation}

The dense fluid with the density $\rho_d$ and the light fluid with the density $\rho_l$ are on the upper and lower sides of the interface, respectively, and fully fill the container. The control parameters in this case are the Atwood number ($At$) and the Reynolds number ($Re$), which are respectively defined as, 

\begin{equation}
A t=\frac{\rho_h-\rho_l}{\rho_h+\rho_l}
\end{equation}
\begin{equation}
\operatorname{Re}=\rho_h \sqrt{L^3 g} / \mu_h
\end{equation}
where $L$ is the length of the domain and $\mu_h$ is the dynamic viscosity of the heavy fluid. In this section, we set $L=1m$, $Re=1024$, $\rho_h=1kg/m^3$, $g=1m/s^2$,  the dynamic viscosity of the light fluid $\mu_l$ is equal to that of the dense fluid $\mu_h$ for all the cases, and the particle size is $dx_0=0.008m~(L/dx_0=125)$ which is the same as that used by Lee and Kim \cite{lee2013numerical}. 

\begin{figure}[H]
\begin{center}
\includegraphics[width=1\textwidth]{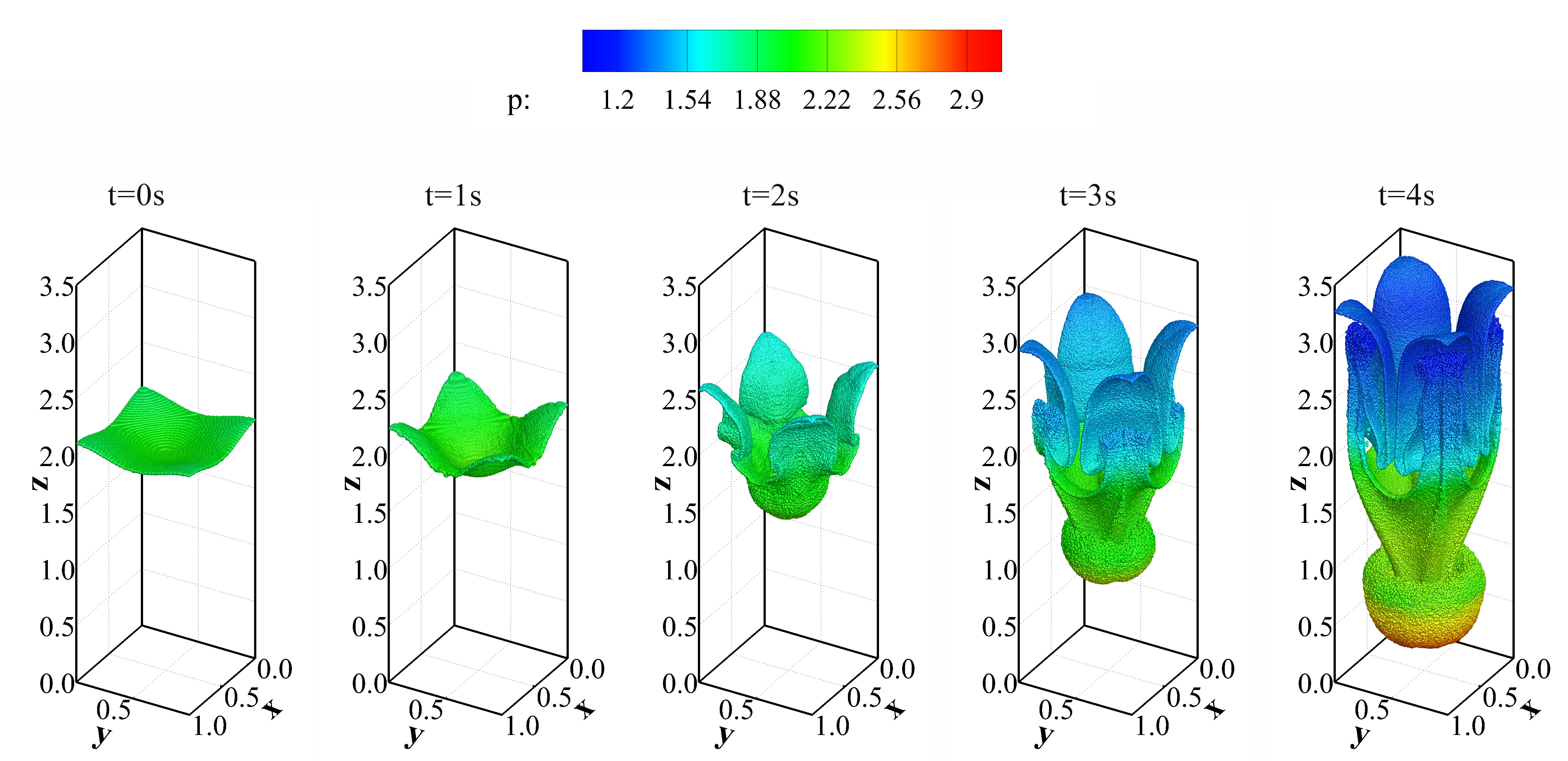}
\caption{Pressure distributions on the interface at $t=0s,~1s,~2s,~3s,~4s$ with $At=0.5$.}\label{fig13}
\end{center}
\end{figure}

Figure 13 illustrates the pressure distributions on the interface at $t=0s,~1s,~2s,~3s,~4s$ with $At=0.5$. With the elapse of time, the originally small interface deformation develops into a very complex shape. The dense fluid sinks down while the light fluid moves up, forming a spherical-like bulge with four petal-shaped tails. The interface pressure is evenly distributed along the direction of gravity.  Figure 14 shows the pressure distributions on the interface with $At=0.2,~0.3,~0.4,~0.5$ at $t=4s$. For the same time instant, the heavy fluid penetrates deeper into the light fluid with increasing $At$, as the density ratio of dense and light fluids increases with $At$. The results observed in the present SPH model are in good agreement with those in the literature \cite{he1999three, lee2013numerical, hamzehloo2021direct}. 

\begin{figure}
\begin{center}
\includegraphics[width=1\textwidth]{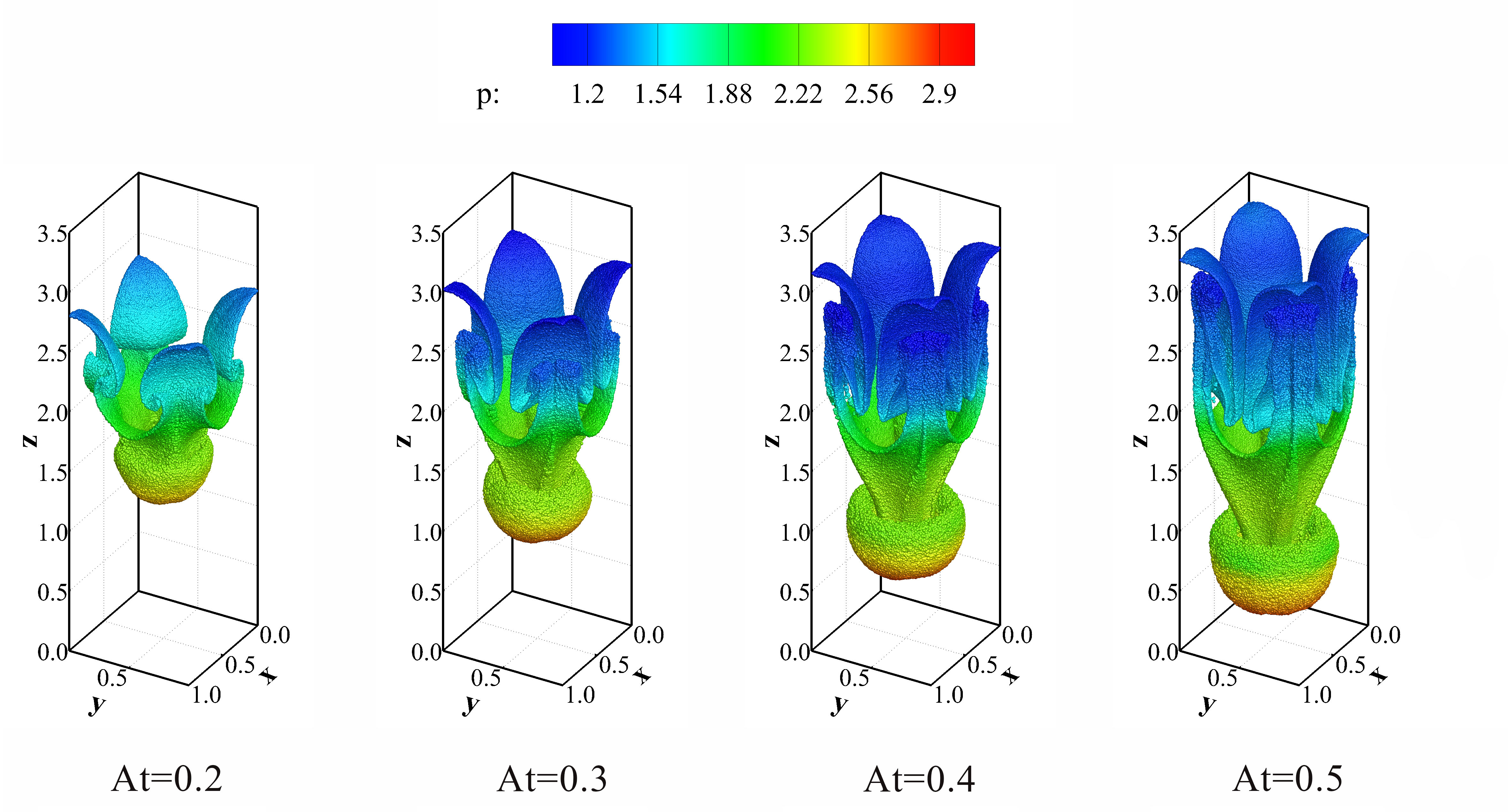}
\caption{Pressure distributions on the interface for $At=0.2,~0.3,~0.4,~0.5$ at $t=4s$.}\label{fig14}
\end{center}
\end{figure}

We also record the lowest position of the spherical-like bulge during the development of the Rayleigh-Taylor instability. The time histories of the lowest position for different $At$  obtained by the present SPH method and LBM of Lee and Kim \cite{lee2013numerical} are shown in Fig. 15. The two numerical results show great agreement, once again verifying the accuracy of the present SPH method.

\begin{figure}[H]
\begin{center}
\includegraphics[width=1\textwidth]{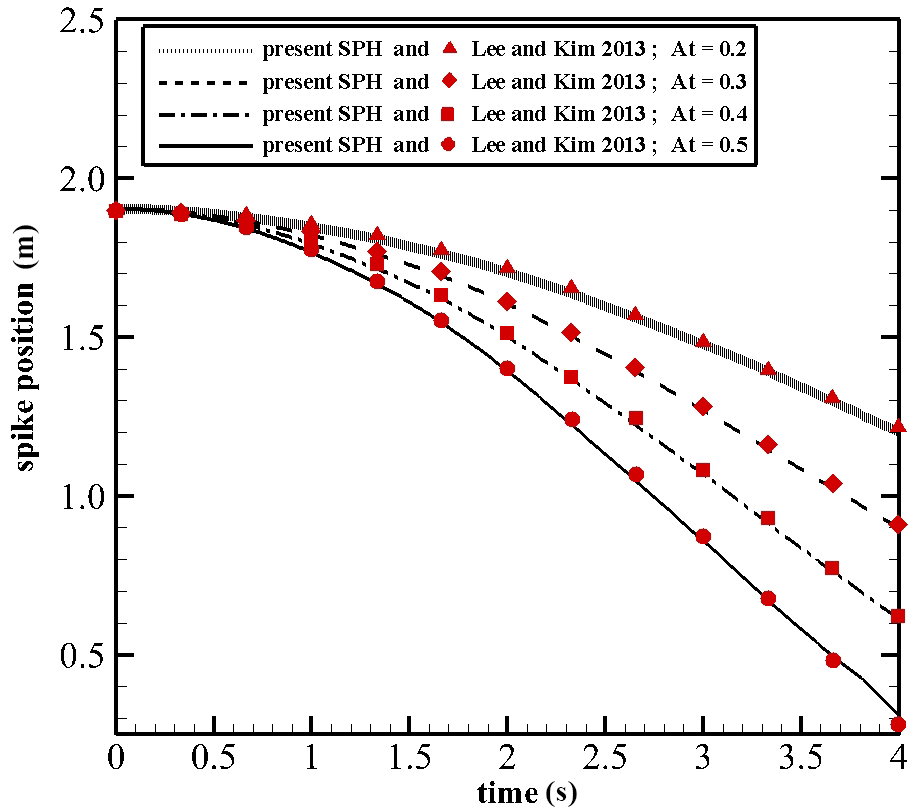}
\caption{Time history of the the lowest position of the spherical-like bulge for $At=0.2,~0.3,~0.4,~0.5$ obtained by the present SPH method and LBM of Lee and Kim \cite{lee2013numerical}.}\label{fig15}
\end{center}
\end{figure}

\subsection{Kelvin-Helmholtz instability}

Finally, the Kelvin-Helmholtz instability is investigated. This problem has been simulated by different SPH methods \cite{chen2015sph,meng2020multiphase}. Here we present the results with high resolution to show more details in the flow field. 

\begin{figure}[H]
\begin{center}
\includegraphics[width=1\textwidth]{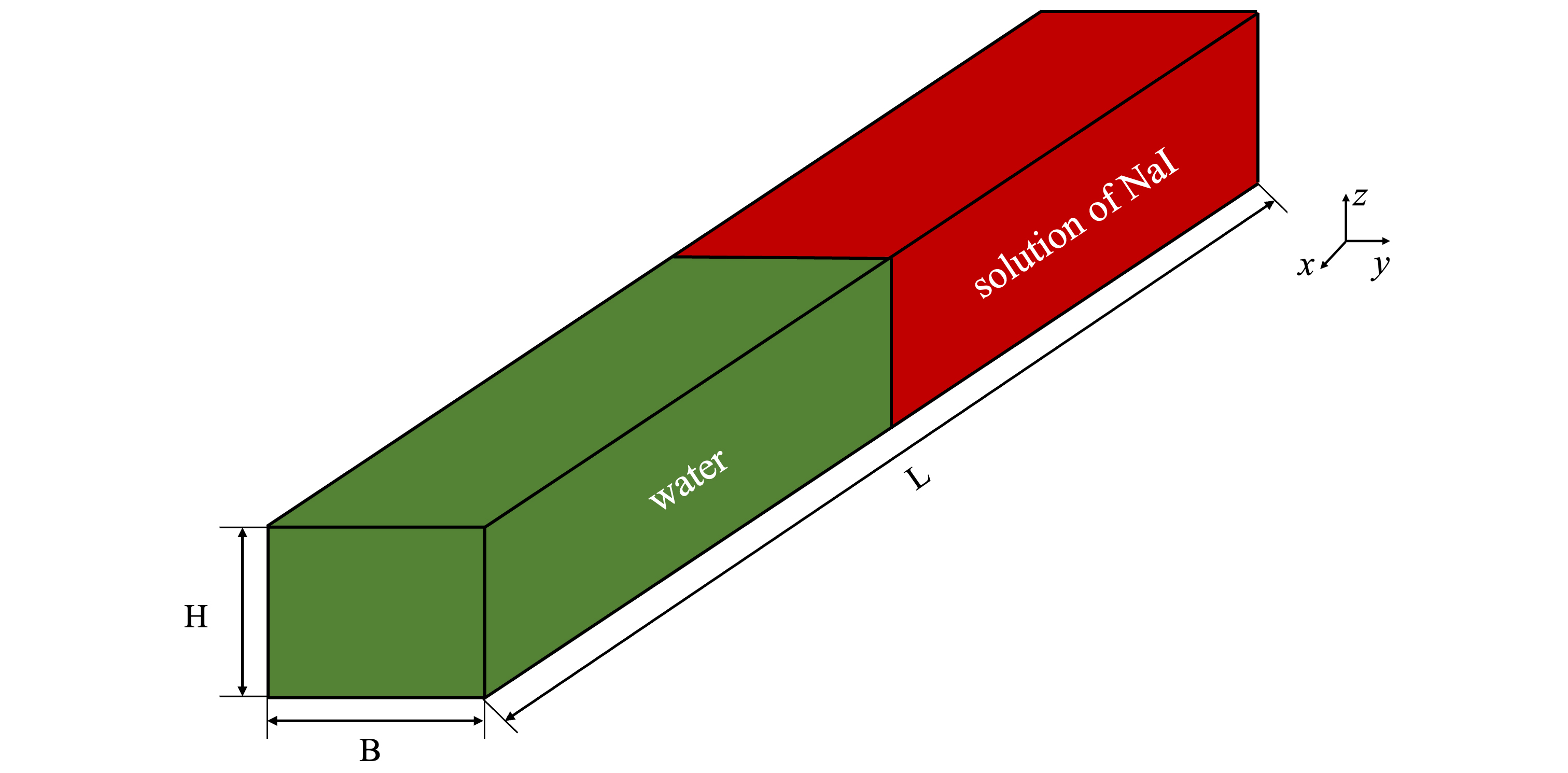}
\caption{Scheme of the initial setup for Kelvin-Helmholtz instability problem.}\label{fig16}
\end{center}
\end{figure}

The initial setup for this case is based on the experimental study on non-Boussinesq gravity currents, which was carried out by Lowe et al. \cite{lowe2005non}. A cuboid fluid domain $\Omega=x \times y \times z=(0,1.81m) \times(-0.1m,0.1m) \times(0,0.2m)$ is filled with the water of the density $\rho_w=998.6 kg/m^3$  in the left half and the sodium iodide (NaI) solution of the density $\rho_{s}=1466.3 kg/m^3$ in the right half. Both of the liquids have the dynamic viscosity of $1\times10^{-3} Pa \cdot s^{-1}$. The gravity is $g=9.81m/s^2$. Due to the difference in the initial densities, the two fluids penetrate each other. The particle size $dx_0=0.0025m~(H/dx_0=80)$ is adopted to simulate this example, which is the same as the smallest particle size in the 2D case of Meng et al. \cite{meng2020multiphase}. 

\begin{figure}
\begin{center}
\includegraphics[width=1\textwidth]{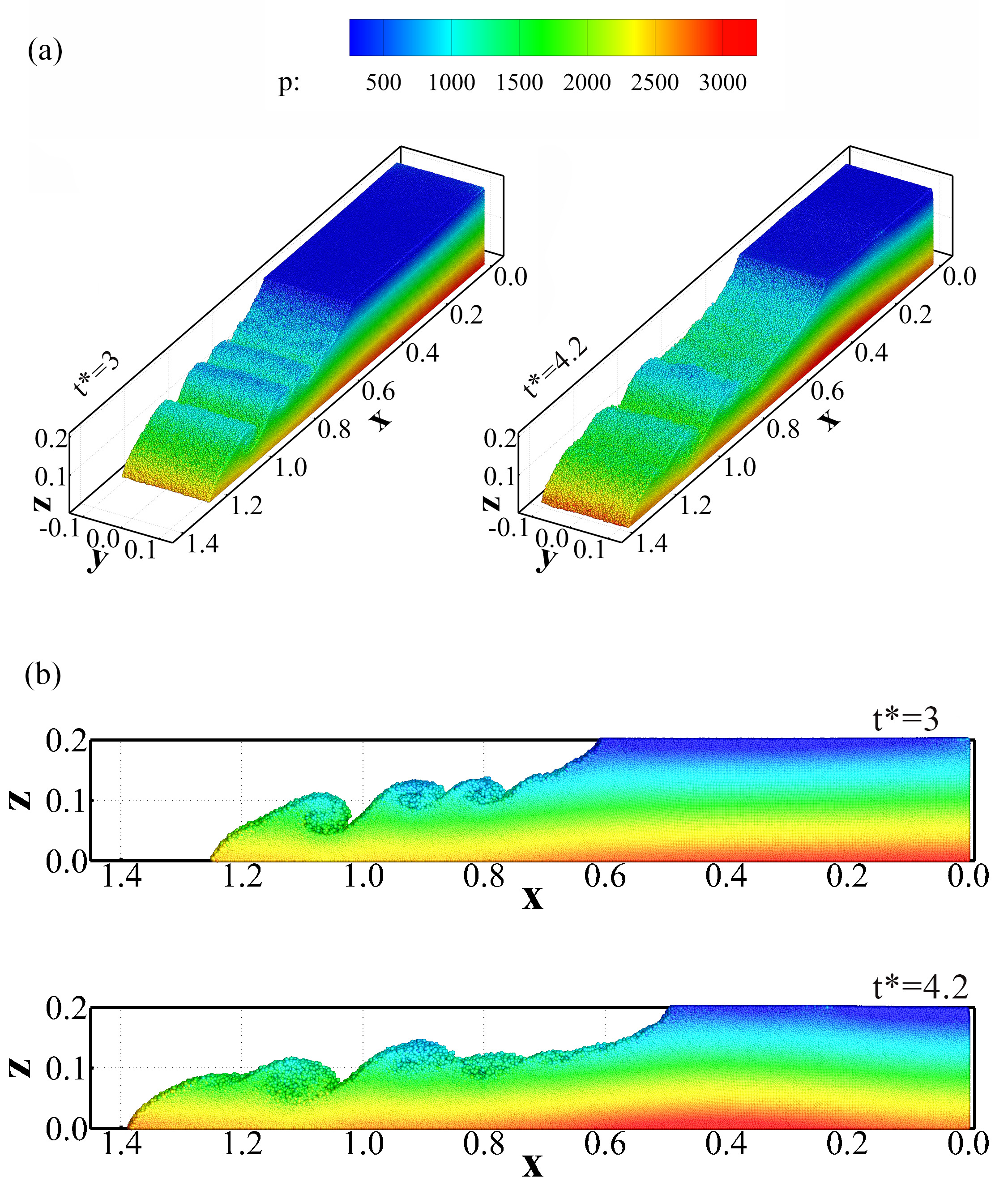}
\caption{Pressure fields of the NaI solution at $t^*=3$ and $t^*=4.2$: (a) 3D view and (b) 2D view along the $-y$ direction.}\label{fig17}
\end{center}
\end{figure}

For a better observation of the shape of the interface, only the phase of the NaI solution is plotted. Figure 17 shows the pressure fields of the NaI solution at $t^*=3$ and $t^*=4.2$, with $t^*=t\sqrt{(g(1-\frac{\rho_{w}}{\rho_{s}}) / H)}$ being the dimensionless time. A very smooth pressure field without non-physical numerical noise can be observed, which demonstrates the ability of the present generalized dissipation term to suppress pressure oscillations. The interface exhibits a pronounced Kelvin-Helmholtz instability, which can be more clearly observed in the 2D view along the $-y$ direction in Fig. 17 (b). Such an intelligible phenomenon of Kelvin-Helmholtz instability was not observed in the previous SPH studies even simulated with the same particle resolution \cite{meng2020multiphase}.

\begin{figure}
\begin{center}
\includegraphics[width=1\textwidth]{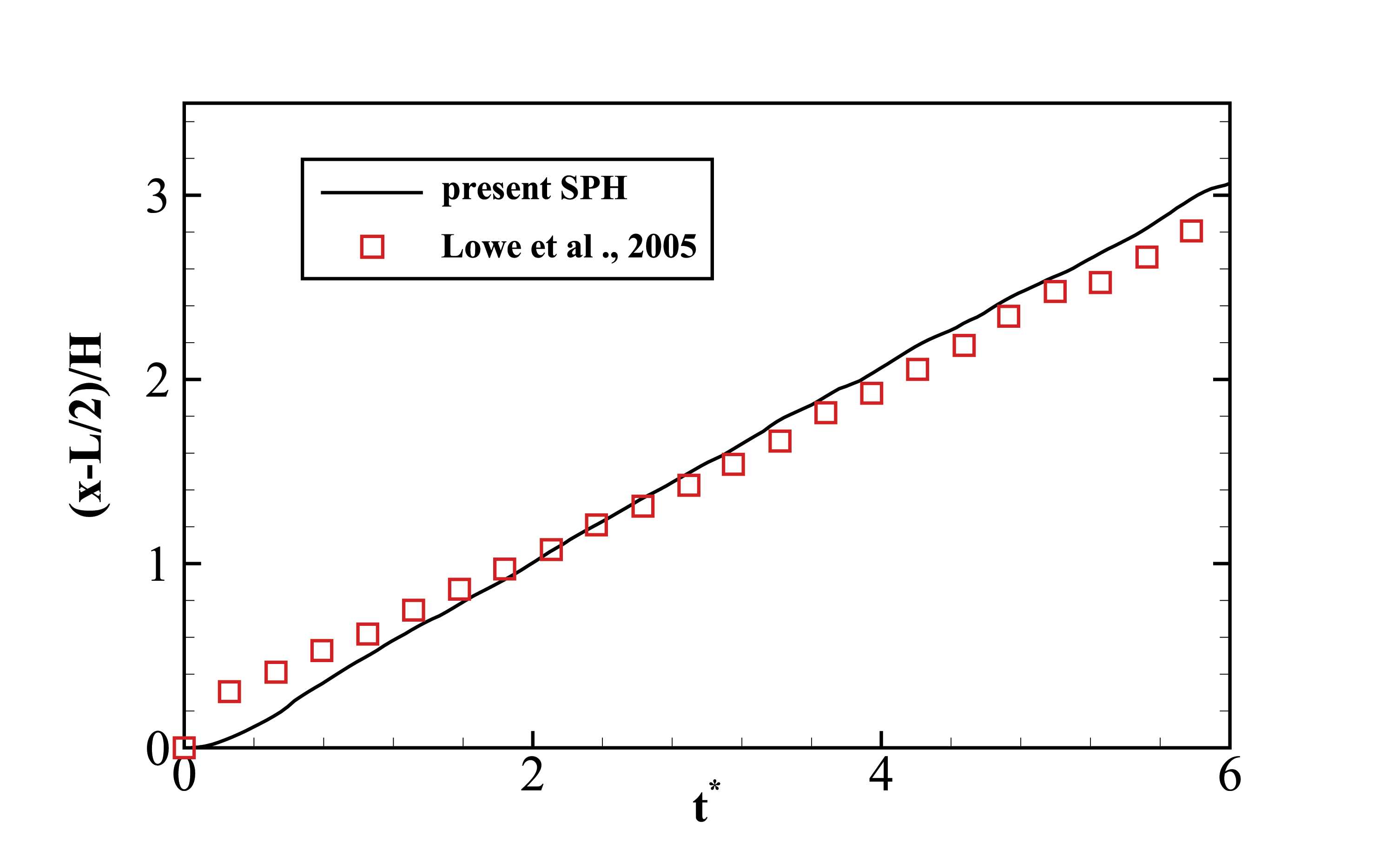}
\caption{Variation of the horizontal position of the water front with time; black line: the numerical results obtained by the present SPH method; red dots: the experimental data of Lowe et al. \cite{lowe2005non}.}\label{fig18}
\end{center}
\end{figure}

Figure 18 plots the variation of the horizontal position of the water front with time obtained by the present SPH method and the corresponding experimental data from Lowe et al. \cite{lowe2005non}. The good agreement between the present results and the experiment data quantitatively validates the accuracy of the present SPH method.

\section{Conclusion}

We propose a generalized density dissipation scheme for weakly-compressible SPH to effectively suppress pressure oscillations in both multiphase and free-surface flow simulations. The need for this improvement arises from that the previous dissipation schemes are unable to cross the interfaces where density is discontinuous. To address this, we introduce an improved dissipation term based on the Laplacian of the density increment. This avoids the problem of undefined Laplacian of density at the interface. The dissipation volume conservation is also considered to ensure a uniform particle distribution in the support domain, especially for large-density-ratio flow simulation. This prompts us to introduce a volume correction factor into the dissipation term. Finally, a generalized density dissipation term suitable for both free-surface and multiphase flow simulations is formed. 

The relationship between the present generalized density dissipation and the basic dissipation in $\delta$-SPH, and the relationship between the present generalized density dissipation and the dissipation in the SPH model with approximate Riemann solver, are discussed. We demonstrate that the present generalized dissipation term recovers the dissipation term in $\delta$-SPH when simulating single-phase flow. Furthermore, we show that the generalized dissipation term and the dissipation produced by SPH with an approximate Riemann solver are essentially the same. This provides a clear explanation on why the approximate Riemann solver can effectively suppress pressure oscillations.

Four numerical examples, i.e., liquid sloshing in a rectangular tank, single and double bubbles rising, Rayleigh-Taylor instability, and Kelvin-Helmholtz instability, are simulated to evaluate the ability of the proposed generalized density dissipation to suppress pressure oscillations. Comparisons between the results obtained by the proposed method and those published in the literature demonstrate its accuracy, stability, and robustness in simulating complex 3D problems involving large density ratio, complex interface, and free-surface with large-deformation. These promising results highlight the potential of the proposed method as a valuable tool in tackling challenging engineering problems.


\section*{Declaration of conflict of interest}

The authors declare that they have no financial and personal relationships with other people or organizations that can inappropriately influence our work and there is no professional or other personal interest of any nature or kind in any product, service or company that could be construed as influencing the position presented in, or the review of, the paper entitled, ``A generalized Density Dissipation for weakly-compressible SPH".

\section*{Acknowledgement}
BXZ and PY gratefully acknowledges financial support from the National Natural Science Foundation of China (Grant No. 12071367), Guangdong Provincial Key Laboratory of Turbulence Research and Applications (Grant No. 2019B21203001) . BXZ and TSC gratefully acknowledges financial support from the Research Council of Norway (Project No. 315110). The authors also thank the computational resources provided  by Center for Computational Science and Engineering, Southern University of Science and Technology.



  \bibliographystyle{elsarticle-num} 
  \bibliography{Boxue_SPH}

\begin{thebibliography}{10}
\expandafter\ifx\csname url\endcsname\relax
  \def\url#1{\texttt{#1}}\fi
\expandafter\ifx\csname urlprefix\endcsname\relax\def\urlprefix{URL }\fi
\expandafter\ifx\csname href\endcsname\relax
  \def\href#1#2{#2} \def\path#1{#1}\fi

\bibitem{liu2019smoothed}
M.~Liu, Z.~Zhang, Smoothed particle hydrodynamics ({SPH}) for modeling
  fluid-structure interactions, Science China Physics, Mechanics \& Astronomy
  62 (2019) 1--38.

\bibitem{sun2021accurate}
P.-N. Sun, D.~Le~Touze, G.~Oger, A.-M. Zhang, An accurate {FSI-SPH} modeling of
  challenging fluid-structure interaction problems in two and three dimensions,
  Ocean Engineering 221 (2021) 108552.

\bibitem{zhang2021multi}
C.~Zhang, M.~Rezavand, X.~Hu, A multi-resolution method for fluid-structure
  interactions, Journal of Computational Physics 429 (2021) 110028.

\bibitem{zhang2017smoothed}
A.-M. Zhang, P.-N. Sun, F.-R. Ming, A.~Colagrossi, Smoothed particle
  hydrodynamics and its applications in fluid-structure interactions, Journal
  of Hydrodynamics, Ser. B 29~(2) (2017) 187--216.

\bibitem{monaghan1994simulating}
J.~J. Monaghan, Simulating free surface flows with {SPH}, Journal of
  Computational Physics 110~(2) (1994) 399--406.

\bibitem{colagrossi2003numerical}
A.~Colagrossi, M.~Landrini, Numerical simulation of interfacial flows by
  smoothed particle hydrodynamics, Journal of Computational Physics 191~(2)
  (2003) 448--475.

\bibitem{HU2006844}
X.~Hu, N.~Adams, A multi-phase {SPH} method for macroscopic and mesoscopic
  flows, Journal of Computational Physics 213~(2) (2006) 844--861.

\bibitem{chen2015sph}
Z.~Chen, Z.~Zong, M.~Liu, L.~Zou, H.~Li, C.~Shu, An {SPH} model for multiphase
  flows with complex interfaces and large density differences, Journal of
  Computational Physics 283 (2015) 169--188.

\bibitem{xu2020modeling}
X.~Xu, M.~Dey, M.~Qiu, J.~J. Feng, Modeling of van der waals force with
  smoothed particle hydrodynamics: Application to the rupture of thin liquid
  films, Applied Mathematical Modelling 83 (2020) 719--735.

\bibitem{monaghan2002sph}
J.~J. Monaghan, {SPH} compressible turbulence, Monthly Notices of the Royal
  Astronomical Society 335~(3) (2002) 843--852.

\bibitem{antuono2021smoothed}
M.~Antuono, S.~Marrone, A.~Di~Mascio, A.~Colagrossi, Smoothed particle
  hydrodynamics method from a large eddy simulation perspective. generalization
  to a quasi-lagrangian model, Physics of Fluids 33~(1) (2021) 015102.

\bibitem{shao2006incompressible}
S.~Shao, Incompressible {SPH} simulation of wave breaking and overtopping with
  turbulence modelling, International Journal for Numerical Methods in Fluids
  50~(5) (2006) 597--621.

\bibitem{hughes2010comparison}
J.~P. Hughes, D.~I. Graham, Comparison of incompressible and
  weakly-compressible {SPH} models for free-surface water flows, Journal of
  Hydraulic Research 48~(sup1) (2010) 105--117.

\bibitem{antuono2012numerical}
M.~Antuono, A.~Colagrossi, S.~Marrone, Numerical diffusive terms in
  weakly-compressible {SPH} schemes, Computer Physics Communications 183~(12)
  (2012) 2570--2580.

\bibitem{dilts1999moving}
G.~A. Dilts, Moving-least-squares-particle hydrodynamics - {I. Consistency} and
  stability, International Journal for Numerical Methods in Engineering 44~(8)
  (1999) 1115--1155.

\bibitem{molteni2009simple}
D.~Molteni, A.~Colagrossi, A simple procedure to improve the pressure
  evaluation in hydrodynamic context using the {SPH}, Computer Physics
  Communications 180~(6) (2009) 861--872.

\bibitem{antuono2010free}
M.~Antuono, A.~Colagrossi, S.~Marrone, D.~Molteni, Free-surface flows solved by
  means of {SPH} schemes with numerical diffusive terms, Computer Physics
  Communications 181~(3) (2010) 532--549.

\bibitem{sun2017deltaplus}
P.~Sun, A.~Colagrossi, S.~Marrone, A.~Zhang, The $\delta$plus-{SPH} model:
  Simple procedures for a further improvement of the {SPH} scheme, Computer
  Methods in Applied Mechanics and Engineering 315 (2017) 25--49.

\bibitem{sun2019consistent}
P.~Sun, A.~Colagrossi, S.~Marrone, M.~Antuono, A.-M. Zhang, A consistent
  approach to particle shifting in the $\delta$-plus-{SPH} model, Computer
  Methods in Applied Mechanics and Engineering 348 (2019) 912--934.

\bibitem{mokos2017multi}
A.~Mokos, B.~D. Rogers, P.~K. Stansby, A multi-phase particle shifting
  algorithm for {SPH} simulations of violent hydrodynamics with a large number
  of particles, Journal of Hydraulic Research 55~(2) (2017) 143--162.

\bibitem{hammani2020detailed}
I.~Hammani, S.~Marrone, A.~Colagrossi, G.~Oger, D.~Le~Touz{\'e}, Detailed study
  on the extension of the $\delta$-{SPH} model to multi-phase flow, Computer
  Methods in Applied Mechanics and Engineering 368 (2020) 113189.

\bibitem{monaghan1997sph}
J.~J. Monaghan, {SPH } and {Riemann} solvers, Journal of Computational Physics
  136~(2) (1997) 298--307.

\bibitem{puri2014approximate}
K.~Puri, P.~Ramachandran, Approximate {Riemann} solvers for the godunov {SPH}
  {(GSPH)}, Journal of Computational Physics 270 (2014) 432--458.

\bibitem{zhang2017weakly}
C.~Zhang, X.~Hu, N.~A. Adams, A weakly compressible {SPH} method based on a
  low-dissipation {Riemann} solver, Journal of Computational Physics 335 (2017)
  605--620.

\bibitem{meng2020multiphase}
Z.-F. Meng, P.-P. Wang, A.-M. Zhang, F.-R. Ming, P.-N. Sun, A multiphase {SPH}
  model based on roe’s approximate {Riemann} solver for hydraulic flows with
  complex interface, Computer Methods in Applied Mechanics and Engineering 365
  (2020) 112999.

\bibitem{meng2021numerical}
Z.-F. Meng, F.-R. Ming, P.-P. Wang, A.~Zhang, Numerical simulation of water
  entry problems considering air effect using a multiphase riemann-sph model,
  Advances in Aerodynamics 3~(1) (2021) 1--16.

\bibitem{fang2022accurate}
X.-L. Fang, A.~Colagrossi, P.-P. Wang, A.-M. Zhang, An accurate and robust
  axisymmetric {SPH} method based on {Riemann} solver with applications in
  ocean engineering, Ocean Engineering 244 (2022) 110369.

\bibitem{zhang2015sph}
A.~Zhang, P.~Sun, F.~Ming, An {SPH} modeling of bubble rising and coalescing in
  three dimensions, Computer Methods in Applied Mechanics and Engineering 294
  (2015) 189--209.

\bibitem{brackbill1992continuum}
J.~U. Brackbill, D.~B. Kothe, C.~Zemach, A continuum method for modeling
  surface tension, Journal of Computational Physics 100~(2) (1992) 335--354.

\bibitem{rezavand2020weakly}
M.~Rezavand, C.~Zhang, X.~Hu, A weakly compressible {SPH} method for violent
  multi-phase flows with high density ratio, Journal of Computational Physics
  402 (2020) 109092.

\bibitem{monaghan2013simple}
J.~J. Monaghan, A.~Rafiee, A simple {SPH} algorithm for multi-fluid flow with
  high density ratios, International Journal for Numerical Methods in Fluids
  71~(5) (2013) 537--561.

\bibitem{grenier2009hamiltonian}
N.~Grenier, M.~Antuono, A.~Colagrossi, D.~Le~Touz{\'e}, B.~Alessandrini, An
  hamiltonian interface {SPH} formulation for multi-fluid and free surface
  flows, Journal of Computational Physics 228~(22) (2009) 8380--8393.

\bibitem{hu2007incompressible}
X.~Hu, N.~A. Adams, An incompressible multi-phase {SPH} method, Journal of
  Computational Physics 227~(1) (2007) 264--278.

\bibitem{flekkoy2000foundations}
E.~G. Flekk{\o}y, P.~V. Coveney, G.~De~Fabritiis, Foundations of dissipative
  particle dynamics, Physical Review E 62~(2) (2000) 2140.

\bibitem{adami2010new}
S.~Adami, X.~Hu, N.~A. Adams, A new surface-tension formulation for multi-phase
  {SPH} using a reproducing divergence approximation, Journal of Computational
  Physics 229~(13) (2010) 5011--5021.

\bibitem{monaghan1989problem}
J.~Monaghan, On the problem of penetration in particle methods, Journal of
  Computational physics 82~(1) (1989) 1--15.

\bibitem{marrone2011delta}
S.~Marrone, M.~Antuono, A.~Colagrossi, G.~Colicchio, D.~Le~Touz{\'e},
  G.~Graziani, $\delta$-{SPH} model for simulating violent impact flows,
  Computer Methods in Applied Mechanics and Engineering 200~(13-16) (2011)
  1526--1542.

\bibitem{flebbe1994smoothed}
O.~Flebbe, S.~Muenzel, H.~Herold, H.~Riffert, H.~Ruder, Smoothed particle
  hydrodynamics: physical viscosity and the simulation of accretion disks, The
  Astrophysical Journal, vol. 431, no. 2, pt. 1, p. 754-760 431 (1994)
  754--760.

\bibitem{takeda1994numerical}
H.~Takeda, S.~M. Miyama, M.~Sekiya, Numerical simulation of viscous flow by
  smoothed particle hydrodynamics, Progress of Theoretical Physics 92~(5)
  (1994) 939--960.

\bibitem{brookshaw1985method}
L.~Brookshaw, A method of calculating radiative heat diffusion in particle
  simulations, Publications of the Astronomical Society of Australia 6~(2)
  (1985) 207--210.

\bibitem{basa2009robustness}
M.~Basa, N.~J. Quinlan, M.~Lastiwka, Robustness and accuracy of {SPH}
  formulations for viscous flow, International Journal for Numerical Methods in
  Fluids 60~(10) (2009) 1127--1148.

\bibitem{fatehi2011error}
R.~Fatehi, M.~T. Manzari, Error estimation in smoothed particle hydrodynamics
  and a new scheme for second derivatives, Computers \& Mathematics with
  Applications 61~(2) (2011) 482--498.

\bibitem{sun2021accurate1}
P.-N. Sun, D.~Le~Touz{\'e}, G.~Oger, A.-M. Zhang, An accurate {SPH} volume
  adaptive scheme for modeling strongly-compressible multiphase flows. part 1:
  Numerical scheme and validations with basic 1d and 2d benchmarks, Journal of
  Computational Physics 426 (2021) 109937.

\bibitem{sun2021accurate2}
P.-N. Sun, D.~Le~Touz{\'e}, G.~Oger, A.-M. Zhang, An accurate {SPH} volume
  adaptive scheme for modeling strongly-compressible multiphase flows. part 2:
  Extension of the scheme to cylindrical coordinates and simulations of 3d
  axisymmetric problems with experimental validations, Journal of Computational
  Physics 426 (2021) 109936.

\bibitem{zheng2019multiphase}
B.~Zheng, Z.~Chen, A multiphase smoothed particle hydrodynamics model with
  lower numerical diffusion, Journal of Computational Physics 382 (2019)
  177--201.

\bibitem{vacondio2021grand}
R.~Vacondio, C.~Altomare, M.~De~Leffe, X.~Hu, D.~Le~Touz{\'e}, S.~Lind, J.-C.
  Marongiu, S.~Marrone, B.~D. Rogers, A.~Souto-Iglesias, Grand challenges for
  smoothed particle hydrodynamics numerical schemes, Computational Particle
  Mechanics 8 (2021) 575--588.

\bibitem{green2019smoothed}
M.~D. Green, R.~Vacondio, J.~Peir{\'o}, A smoothed particle hydrodynamics
  numerical scheme with a consistent diffusion term for the continuity
  equation, Computers \& Fluids 179 (2019) 632--644.

\bibitem{xu2016improved}
X.~Xu, An improved {SPH} approach for simulating 3d dam-break flows with
  breaking waves, Computer methods in applied Mechanics and Engineering 311
  (2016) 723--742.

\bibitem{wang2021new}
P.-P. Wang, A.-M. Zhang, Z.-F. Meng, F.-R. Ming, X.-L. Fang, A new type of weno
  scheme in {SPH} for compressible flows with discontinuities, Computer Methods
  in Applied Mechanics and Engineering 381 (2021) 113770.

\bibitem{doring2005developpement}
M.~Doring, D{\'e}veloppement d'une m{\'e}thode {SPH} pour les applications
  {\`a} surface libre en hydrodynamique, Ph.D. thesis, Nantes (2005).

\bibitem{zheng2021novel}
B.~Zheng, L.~Sun, P.~Yu, A novel interface method for two-dimensional
  multiphase {SPH}: Interface detection and surface tension formulation,
  Journal of Computational Physics 431 (2021) 110119.

\bibitem{liu2009three}
D.~Liu, P.~Lin, Three-dimensional liquid sloshing in a tank with baffles, Ocean
  Engineering 36~(2) (2009) 202--212.

\bibitem{molin2012experimental}
B.~Molin, F.~Remy, C.~Audiffren, R.~Marcer, Experimental and numerical study of
  liquid sloshing in a rectangular tank with three fluids, in: The
  Twenty-second International Offshore and Polar Engineering Conference,
  OnePetro, 2012.

\bibitem{eymard2000finite}
R.~Eymard, T.~Gallou{\"e}t, R.~Herbin, Finite volume methods, Handbook of
  Numerical Analysis 7 (2000) 713--1018.

\bibitem{hirt1981volume}
C.~W. Hirt, B.~D. Nichols, Volume of fluid {(VOF)} method for the dynamics of
  free boundaries, Journal of Computational Physics 39~(1) (1981) 201--225.

\bibitem{brereton1991coaxial}
G.~Brereton, D.~Korotney, Coaxial and oblique coalescence of two rising
  bubbles, in: {Dynamics of Bubbles and Vortices near a Free Surface, Vol.
  119}, ASME New York, AMD, 1991,  pp.50--73.

\bibitem{sitompul2019filtered}
Y.~P. Sitompul, T.~Aoki, A filtered cumulant lattice {Boltzmann} method for
  violent two-phase flows, Journal of Computational Physics 390 (2019) 93--120.

\bibitem{van2005numerical}
M.~van Sint~Annaland, N.~Deen, J.~Kuipers, Numerical simulation of gas bubbles
  behaviour using a three-dimensional volume of fluid method, Chemical
  Engineering Science 60~(11) (2005) 2999--3011.

\bibitem{yu2016improved}
C.~Yu, Z.~Ye, T.~W. Sheu, Y.~Lin, X.~Zhao, An improved interface preserving
  level set method for simulating three dimensional rising bubble,
  International Journal of Heat and Mass Transfer 103 (2016) 753--772.

\bibitem{yan2020higher}
J.~Yan, S.~Li, X.~Kan, A.-M. Zhang, X.~Lai, Higher-order nonlocal theory of
  {Updated Lagrangian Particle Hydrodynamics (ULPH)} and simulations of
  multiphase flows, Computer Methods in Applied Mechanics and Engineering 368
  (2020) 113176.

\bibitem{he1999three}
X.~He, R.~Zhang, S.~Chen, G.~D. Doolen, On the three-dimensional
  {Rayleigh--Taylor} instability, Physics of Fluids 11~(5) (1999) 1143--1152.

\bibitem{lee2013numerical}
H.~G. Lee, J.~Kim, Numerical simulation of the three-dimensional
  {Rayleigh--Taylor} instability, Computers \& Mathematics with Applications
  66~(8) (2013) 1466--1474.

\bibitem{hamzehloo2021direct}
A.~Hamzehloo, P.~Bartholomew, S.~Laizet, Direct numerical simulations of
  incompressible {Rayleigh--Taylor} instabilities at low and medium atwood
  numbers, Physics of Fluids 33~(5) (2021) 054114.

\bibitem{lowe2005non}
R.~J. Lowe, J.~W. Rottman, P.~Linden, The non-boussinesq lock-exchange problem.
  part 1. theory and experiments, Journal of Fluid Mechanics 537 (2005)
  101--124.

\end{thebibliography}





\end{document}